\documentclass[12pt]{article}

\usepackage{epsf,amsfonts,hyperref}
\renewcommand{\theequation}{\thesection.\arabic{equation}}

\renewcommand{\appendix}[1]{
    \setcounter{section}{0}
    \setcounter{equation}{0}
    \renewcommand{\thesection}{\Alph{section}}
		\renewcommand{\theequation}{\Alph{section}-\arabic{equation}}
}

\newcommand\encadremath[1]{\vbox{\hrule\hbox{\vrule\kern8pt 
\vbox{\kern8pt \hbox{$\displaystyle #1$}\kern8pt} 
\kern8pt\vrule}\hrule}}
\def\enca#1{\vbox{\hrule\hbox{
\vrule\kern8pt\vbox{\kern8pt \hbox{$\displaystyle #1$}
\kern8pt} \kern8pt\vrule}\hrule}}

\newcommand\figureframex[3]{
\begin{figure}[bth]
\hrule\hbox{\vrule\kern8pt 
\vbox{\kern8pt \vbox{
\begin{center}
{\mbox{\epsfxsize=#1.truecm\epsfbox{#2}}}
\end{center}
\caption{#3}
}\kern8pt} 
\kern8pt\vrule}\hrule
\end{figure}
}
\newcommand\figureframey[3]{
\begin{figure}[bth]
\hrule\hbox{\vrule\kern8pt 
\vbox{\kern8pt \vbox{
\begin{center}
{\mbox{\epsfysize=#1.truecm\epsfbox{#2}}}
\end{center}
\caption{#3}
}\kern8pt} 
\kern8pt\vrule}\hrule
\end{figure}
}

\newcommand{\eq}[1]{eq.(\ref{#1})}

\newcommand{\beq}{\begin{equation}}
\newcommand{\eeq}{\end{equation}}
\newcommand{\bea}{\begin{eqnarray}}
\newcommand{\eea}{\end{eqnarray}}

\renewcommand{\thesection}{\arabic{section}}
\renewcommand{\theequation}{\arabic{section}-\arabic{equation}}
\makeatletter
\@addtoreset{equation}{section}
\makeatother
\newtheorem{theorem}{Theorem}[section]
\newtheorem{conjecture}{Conjecture}[section]
\newtheorem{remark}{Remark}[section]
\newtheorem{proposition}{Proposition}[section]
\newtheorem{lemma}{Lemma}[section]
\newtheorem{corollary}{Corollary}[section]
\newtheorem{definition}{Definition}[section]
\def\br{\begin{remark}\rm\small}
\def\er{\end{remark}}
\def\bt{\begin{theorem}}
\def\et{\end{theorem}}
\def\bd{\begin{definition}}
\def\ed{\end{definition}}
\def\bp{\begin{proposition}}
\def\ep{\end{proposition}}
\def\bl{\begin{lemma}}
\def\el{\end{lemma}}
\def\bc{\begin{corollary}}
\def\ec{\end{corollary}}
\def\beaq{\begin{eqnarray}}
\def\eeaq{\end{eqnarray}}
\newcommand{\proof}[1]{{\noindent \bf proof:}\par
{#1} $\square$}

\renewcommand{\and}{{\qquad {\rm and} \qquad}}

\newcommand{\virg}{{\qquad , \qquad}}

 \newcommand{\Tr}{{\,\rm Tr}\:}
\newcommand{\tr}{{\,\rm tr}\:}

\newcommand{\Res}{\mathop{\,\rm Res\,}}

\newcommand{\td}[1]{{\tilde{#1}}}

\renewcommand{\l}{\lambda}
\newcommand{\om}{\omega}

\newcommand{\ee}[1]{{{\rm e}^{#1}}}

\newcommand{\Pint}{{\int\kern -1.em -\kern-.25em}}

\renewcommand{\l}{\lambda}

\newcommand{\Kerec}{{\mathbb K}}

\begin{document}
\sloppy


\pagestyle{empty}
\hfill SPT-09/015
\addtolength{\baselineskip}{0.20\baselineskip}
\begin{center}
\vspace{26pt}
{\large \bf {Determinantal formulae and loop equations}}
\newline
\vspace{26pt}

{\sl M.\ Berg\`ere}\hspace*{0.05cm}\footnote{ E-mail: bergere@spht.saclay.cea.fr },
{\sl B.\ Eynard}\hspace*{0.05cm}\footnote{ E-mail: eynard@spht.saclay.cea.fr }\\
\vspace{6pt}

Institut de Physique Th\'eorique, CEA, IPhT,\\
F-91191 Gif-sur-Yvette, France,\\
CNRS, URA 2306, F-91191 Gif-sur-Yvette, France.\\
\end{center}

\vspace{20pt}
\begin{center}
{\bf Abstract}

We prove that the correlations functions, generated by the determinantal process of the Christoffel-Darboux kernel of an arbitrary order $2$ ODE, do satisfy loop equations.

\end{center}
%





\vspace{26pt}
\pagestyle{plain}
\setcounter{page}{1}



\section{Introduction}

It is well known that matrix models satisfy both loop equations \cite{ZJDFG}, and determinantal formulae \cite{Mehtabook, eynmehta, borsosh, soshnikov, harnad1}.
However, both notions (loop equations and determinantal formulae) exist beyond matrix models.
In this paper we show that the correlators obtained from determinants of the Christoffel-Darboux kernel of an arbitrary differential system of order 2, do satisfy loop equations.

\bigskip

For matrix models, the "fermionic" correlators are expectation values of ratios of characteristic polynomials:
\beq
{\cal K}_n(x_1,\dots,x_n;y_1,\dots,y_n) = \left< {\prod_{i=1}^n \det(x_i-M)\over \prod_{i=1}^n \det(y_i-M)}\right>
\eeq
where $<.>$ is the expectation value with some matrix measure $dM\,\ee{-\Tr V(M)}$.
It was found \cite{FyoStr, bergere, AkV, AkP} that there exists some kernels ${\cal K}_{i,j}$, such that these correlators satisfy Giambelli-type  determinantal formulae:
\beq\label{calKndetcalK}
{\cal K}_n(x_1,\dots,x_n;y_1,\dots,y_n) = {\prod_{i,j} (x_i-y_j)\over \prod_{i>j} (x_i-x_j)\,\,\prod_{i>j} (y_i-y_j)}\,\,{\det({\cal K}_{i,j}(x_i;y_j))}
\eeq

\medskip

On the other hand the "bosonic" correlators are expectation values of product of traces:
\beq
{\cal W}_n(x_1,\dots,x_n) = \left< \prod_{i=1}^n \Tr{1\over x_i-M} \right>
\eeq
and in fact we prefer to consider their cumulants (also called connected parts):
\beq
W_n(x_1,\dots,x_n) = \left< \prod_{i=1}^n \Tr{1\over x_i-M} \right>_c
\eeq

It is clear that traces can be obtained as some limits of determinants:
\beq
\mathop{{\rm lim}}_{y\to x} {\partial \over \partial x}\,{\det(x-M)\over \det(y-M)}  = \Tr {1\over x-M}
\eeq
and one finds the determinantal formulae for the ${\cal W}_n$'s:
\beq
{\cal W}_n(x_1,\dots,x_n) = "\det(K(x_i,x_j))"
\eeq
where the bracketed determinant $"\det"$ introduced by \cite{bergere}, consists in renormalizing the 1 and 2 point function (the meaning of this bracket notation is explained below in def. \ref{defWncdetK}), and where $K$ is a kernel related to the kernels ${\cal K}_{i,j}$ in \eq{calKndetcalK}, see def. \ref{defWncdetK} below.

\medskip

In other words, knowing $K$, i.e. knowing fermionic correlators, we can find the $W_n$'s, i.e. bosonic ones.

\medskip

Conversely, one may reconstruct $K$ from the $W_n$'s, under the condition that the $W_n$'s obey some loop equations.
The formula which holds for matrix models, is an exponential formula:
\beq
K(x,y) = \exp{\left(\sum_{n} {1\over n!} \int_y^x\dots\int_y^x W_n\right)}
\eeq
This formula is nothing but a rewriting of the Heine formula \cite{Szego} for matrix models, or also it can be viewed as the Sato formula of integrable systems \cite{BBT, kostovhirota}, it simply follows from 
\beq
{\det(x-M)\over \det{(y-M)}} = \ee{\Tr\big( \ln{(x-M)}-\ln{(y-M)} \big)} = \ee{\int_y^x  \Tr {dx'\over x'-M}}
\eeq

\medskip

Here, we are going to prove that those formulae continue to hold even without an underlying matrix model.

\section{From determinants towards loop equations}
\label{secdettoloop}

In this section, we start from a $2\times 2$ differential system $\Psi' = {\cal D}.\Psi$, we define its Christoffel-Darboux kernel $K$ \cite{Mehtabook}, and
we define the correlation functions $W_n$ from a determinantal formula.
We prove that they satisfy loop equations.

\smallskip

All this section could hopefully be generalized to higher order $m\times m$ differential systems with $m>2$, but for simplicity, we consider only the $2\times 2$ case here.

\smallskip

Again, we emphasize that all the present section is independent from any underlying random matrix problem.

\subsection{Differential system of order $2$}

Consider a differential system of order $2$:
\beq
{d \over d x} \Psi(x) = {\cal D}(x)\,\Psi(x)
\virg
\Psi(x) = \pmatrix{ \psi(x)& \phi(x)   \cr \td\psi(x) & \td\phi(x)}
\virg
\det \Psi =1
\eeq
where ${\cal D}(x)$ is a traceless matrix, whose coefficients $a,b,c,d$ are rational functions of $x$:
\beq
{\cal D}(x) = \pmatrix{ a(x) & b(x) \cr c(x) & d(x)}
\virg
\Tr {\cal D} = a+d =0
\eeq
Notice that we have:
\beq\label{eqDcalpsiphi}
{\cal D} = \Psi' \, \Psi^{-1}=
\pmatrix{
\psi' \td\phi-\phi' \td\psi &  \quad \phi'\psi -\psi'\phi \cr
\td\psi' \td\phi - \td\phi'\td\psi &  \quad \td\phi' \psi -\td\psi' \phi 
}
=\pmatrix{ a & b \cr c & d}
\eeq
Notice that since $\Tr {\cal D}=0$, we have:
\beq\label{dualityD}
{\cal D}^t\, A = - A \, {\cal D}
\virg
A=\pmatrix{0 & 1 \cr -1 & 0}
\eeq

We define the spectral curve as the characteristic polynomial of ${\cal D}$:
\bd
The spectral curve is:
\beq
\hat{\cal E}(x,y) = \det(y-{\cal D}(x))=y^2 - {1\over 2}\tr {\cal D}^2(x) = y^2  + a(x)d(x)-b(x)c(x)
\eeq
\ed
The equation $\hat{\cal E}(x,y)=0$ is an algebraic equation, and it is hyperelliptical since it is of degree $2$ in $y$.

\medskip

Then we define the Christoffel-Darboux kernel associated to $\Psi$:
\bd
We define the Christoffel-Darboux kernel $K(x_1,x_2)$ associated to $\Psi$, by:
\beq
K(x_1,x_2) = {\psi(x_1) \td\phi(x_2) - \td\psi(x_1)\phi(x_2) \over x_1-x_2}
= {1\over x_1-x_2}\,\vec\psi(x_1)^t\,A\, \vec\phi(x_2)
\eeq
where:
\beq
\vec\psi(x_1) = \pmatrix{\psi(x_1) \cr \td\psi(x_1)}
\virg
\vec\phi(x_2) = \pmatrix{\phi(x_2) \cr \td\phi(x_2)}
\virg
A=\pmatrix{0 & 1 \cr -1 & 0}
\eeq
 $A$ is called the Christoffel-Darboux matrix \cite{BEH}.
\ed

Then we define the correlation functions:

\bd\label{defCorrWdetK} We define the {\bf connected} correlation functions by:
\bea
{W}_1(x) 
&=& \mathop{{\rm lim}}_{x'\to x} \left( K(x,x')-{1\over x-x'} \right) = \psi'(x) \td\phi(x) - \td\psi'(x)\phi(x) \cr
&=& \vec\psi(x)^t {\cal D}(x)^t \, A\, \vec\phi(x) = - \vec\psi(x)^t\,A\,{\cal D}(x)\, \vec\phi(x)
\eea
and for $n\geq 2$:
\beq
{W}_n(x_1,\dots,x_n) = 
 - {\delta_{n,2}\over (x_1-x_2)^2} -(-1)^n\, \sum_{\sigma={\rm cyles}} \prod_{i=1}^n K(x_i,x_{\sigma(i)}) 
\eeq
\ed

\bd\label{defWncdetK}
We also define the {\bf non-connected} correlation functions by the determinantal formulae:
\beq
{\cal W}_n(x_1,\dots,x_n) = 
"\det"\left( K(x_i,x_j) \right)
\eeq
where the quotation mark in $"\det"$ means that, when we expand the determinant as a sum of permutations, each time we have a $K(x_i,x_{\sigma(i)})$ with $\sigma(i)=i$ we have to replace it by $W_1(x_i)$, and each time we have the product $K(x_i,x_j) K(x_j,x_i)$, we have to replace it by $K(x_i,x_j) K(x_j,x_i)+{1\over (x_i-x_j)^2}$.
This definition is inspired from \cite{bergere}.

\smallskip
The connected correlation functions are the {\bf cumulants} of the non-connected ones.

\ed


The purpose of these definitions, is that, for a matrix model $\int dM\,\, \ee{-\Tr V(M)}$, it coincides with correlation functions of eigenvalues. In that case, $\psi(x)\,\ee{{1\over 2}V(x)}=p_n(x) $ and $\td\psi(x)\,\ee{{1\over 2}V(x)}=p_{n-1}(x)$ are consecutive orthonormal polynomials of degrees $n$ and $n-1$, orthonormal with respect to the measure $\ee{-V(x)}dx$. And $\phi\,\ee{-{1\over 2}V(x)}=\hat{p}_n$ and $\td\phi\,\ee{-{1\over 2}V(x)}=\hat{p}_{n-1}$ are their Hilbert transforms (see \cite{bergere}):
\beq
\hat{p}_n(x) = \int {dx'\over x-x'}\, p_n(x')\,\, \ee{-V(x')}
\eeq
Thus in the case of matrix models we have:
\beq
K(x_1,x_2) = \,\ee{-{1\over 2}(V(x_1)-V(x_2))}\,\,\, {p_n(x_1)\hat{p}_{n-1}(x_2)-p_{n-1}(x_1)\hat{p}_{n}(x_2)\over x_1-x_2}
\eeq

\subsubsection{Examples}

For instance we have:

$\bullet$ The 1-point function is:
\beq
W_1(x) =  \psi'(x) \td\phi(x) - \td\psi'(x)\phi(x)
\eeq

$\bullet$ The 2-point function is:
\bea\label{W2KKpsi}
&& W_2(x_1,x_2) \cr
&=& -K(x_1,x_2)K(x_2,x_1) - {1\over (x_1-x_2)^2} \cr
&=& -\,{1\over (x_1-x_2)^2}\,
\left( \psi(x_1)\td\psi(x_2)-\td\psi(x_1)\psi(x_2) \right)
\,\left( \phi(x_1)\td\phi(x_2)-\td\phi(x_1)\phi(x_2) \right) \cr
\eea
One can check (see section \ref{secloopeq} below) that:
\beq
{\cal W}_2(x,x)  = W_2(x,x) + W_1(x)^2 = b(x) c(x) - a(x)d(x)  = - \det{{\cal D}(x)}
\eeq
which is a rational function of $x$.

$\bullet$ The 3-point function $W_3$ is:
\beq
W_3(x_1,x_2,x_3) = K(x_1,x_2)K(x_2,x_3)K(x_3,x_1) + K(x_1,x_3)K(x_3,x_2)K(x_2,x_1)
\eeq
it is the cumulant of ${\cal W}_3$:
\bea
{\cal W}_3(x_1,x_2,x_3)
&=& "\det"\pmatrix{ 
K(x_1,x_1) & K(x_1,x_2) & K(x_1,x_3) \cr 
K(x_2,x_1) & K(x_2,x_2) & K(x_2,x_3) \cr
K(x_3,x_1) & K(x_3,x_2) & K(x_3,x_3) \cr}  \cr
&& \cr
&=& \det\pmatrix{ 
W_1(x_1) & K(x_1,x_2) & K(x_1,x_3) \cr 
K(x_2,x_1) & W_1(x_2) & K(x_2,x_3) \cr
K(x_3,x_1) & K(x_3,x_2) & W_1(x_3) \cr} \cr
&& \quad - {W_1(x_1)\over (x_2-x_3)^2}- {W_1(x_2)\over (x_1-x_3)^2}- {W_1(x_3)\over (x_1-x_2)^2}  \cr
\eea

\medskip
and so on...

\subsection{Correlators in terms of a rank 1 matrix $M(x)$}

Using the fact that $K$ is a product of the form:
\beq
K(x_1,x_2) =  {1\over x_1-x_2}\,\vec\psi(x_1)^t\,A\, \vec\phi(x_2)
\eeq
we define the following rank 1 matrix:
\beq
M(x) = \vec\phi(x)\,\vec\psi(x)^t\,A\, 
= \pmatrix{ -\phi(x) \td\psi(x) & \phi(x)\psi(x) \cr - \td\phi(x) \td\psi(x) & \td\phi(x) \psi(x)}
\eeq
$M(x)$ is a rank 1 matrix, and it is a projector on state $\vec\phi(x)$:
\beq
M(x)^2 = M(x)
\virg
M(x) \vec\phi(x) = \vec\phi(x)
\virg
M(x) \vec\psi(x) = 0
\eeq
Its eigenvalues are $1$ and $0$, and in particular:
\beq
\Tr M(x) = 1
\eeq
We have the duality formula:
\beq
A\, M(x)^t\, A = M(x)-1
\eeq

Notice, that thanks to property \eq{dualityD}, $M(x)$ satisfies the Lax equation:
\beq
{\partial_x}\, M(x)= [{\cal D}(x),M(x)]
\eeq

A mere rewriting of the definition \ref{defCorrWdetK} of correlation functions gives:
\bt
\beq
W_1(x) = -\Tr {\cal D}(x) M(x)
\eeq
\beq\label{hatW2MM}
W_2(x_1,x_2) = - {1-\Tr M(x_1)M(x_2)\over (x_1-x_2)^2} = -\, {\Tr (M(x_1)-M(x_2))^2\over 2\,(x_1-x_2)^2}
\eeq
and if $n>2$:
\bea
&& W_n(x_1,\dots,x_n) \cr
&=&  {(-1)^{n+1}}\,  \Tr\, \sum_{\sigma= {\rm cyclic}}\, \prod_i {M(x_{\sigma(i)})\over x_{\sigma(i)}-x_{\sigma(i+1)}} \cr
&=& {(-1)^{n+1}\over n}\,\sum_{\sigma\in S_n}\, {\Tr\left( M(x_{\sigma(1)})\,M(x_{\sigma(2)})\, \dots \, M(x_{\sigma(n)})\, \right)\over (x_{\sigma(1)}-x_{\sigma(2)})\,(x_{\sigma(2)}-x_{\sigma(3)})\,\dots\,\,(x_{\sigma(n)}-x_{\sigma(1)})}  \cr
\eea

\et

For Example:
\beq
W_3(x_1,x_2,x_3) = {\Tr\Big( M(x_1)M(x_2)M(x_3) - M(x_1)M(x_3)M(x_2)\Big)\over  (x_1-x_2)(x_2-x_3)(x_3-x_1)}
\eeq

\subsection{Loop operator}

\bd\label{defloopinsD2}
The Loop operator is a derivation $\delta_x$ such that:
\beq
\delta_{x_2} \psi(x_1) = - K(x_1,x_2) \psi(x_2)
\virg
\delta_{x_2} \td\psi(x_1) = - K(x_1,x_2) \td\psi(x_2)
\eeq
\beq
\delta_{x_2} \phi(x_1) = - K(x_2,x_1) \phi(x_2)
\virg
\delta_{x_2} \td\phi(x_1) = - K(x_2,x_1) \td\phi(x_2)
\eeq
and $\delta_x (u v) = v\, \delta_x u + u\,\delta_x v$, and ${d\over dx_1} \delta_{x_2} = \delta_{x_2}{d\over dx_1} $.
\ed

In the matrix models litterature, it coincides with the "loop insertion operator" \cite{ZJDFG, Ak, Amb}, often denoted $\delta_x = {\partial \over \partial V(x)}$.

\subsubsection{Some properties of loop operators}

The definitions of Def.\ref{defloopinsD2} can be rewritten as:
\bt\label{thdeltapsiphi}
\beq
\delta_y \vec\psi(x)  = {(1-M(y))\over y-x} \, \vec\psi(x)
\virg
\delta_y \vec\psi(x)^t\, A =  \vec\psi(x)^t\,A\,{M(y)\over y-x}
\eeq
and:
\beq
\delta_y \vec\phi(x) 
= {M(y)\,\vec\phi(x)\over x-y} 
\virg
\delta_y \vec\phi(x)^t A =  \vec\phi(x)^t A\,{1-M(y)\over x-y}
\eeq
\et

From which it is easy to derive:
\bt
We have:
\beq
\delta_y\,M(x) = {1\over (x-y)}\,[M(y),M(x)]=\delta_x\,M(y)
\eeq
\et
And:
\bt
We have:
\beq
\delta_{x_3} K(x_1,x_2) =  - K(x_1,x_3)\,K(x_3,x_2)
\eeq
\et

From this last theorem, one easily finds:
\bt
We have:
\beq
 \delta_{x_{n+1}} W_n(x_1,\dots,x_n) =  W_{n+1}(x_1,\dots,x_n,x_{n+1}) + {\delta_{n,1}\over (x_1-x_2)^2}
\eeq
\et

Then, we have the following property, which is very useful:
\bt\label{thdeltaD}
We have:
\beq
\delta_y\,{\cal D}(x) = {1\over (x-y)^2}\,({\rm Id}-M(x)-M(y))+{1\over x-y}\,[M(y),{\cal D}(x)]
\eeq
\et
\proof{
This theorem is quite technical and the proof is given in appendix \ref{appthdeltaD}.
The proof consists in applying $\delta_y$ to:
\beq
\vec\phi'(x) = {\cal D}(x)\,\vec\phi(x)
\eeq
}

\subsection{Loop equations}
\label{secloopeq}

In this section, we prove that the correlation functions $W_n$ satisfy a set of "loop equations".

\bigskip

In that purpose we need to introduce some definitions. 

\bd Definition of the rational functions $P_n$:

\beq
P_1(x) = \det {\cal D}(x) = -\,{1\over 2}\,\Tr {\cal D}^2(x)
\eeq
\beq
P_2(x;x_2) =  \Tr{{\cal D}(x)-{\cal D}(x_2)-(x-x_2){\cal D}'(x_2)\over (x-x_2)^2}\, M(x_2)
\eeq
and if $n\geq 2$:
\beq
Q_{n+1}(x;x_1,\dots,x_n) = \sum_\sigma {\Tr {\cal D}(x)\,M(x_{\sigma(1)})M(x_{\sigma(2)})\dots M(x_{\sigma(n)}) \over (x-x_{\sigma(1)})(x_{\sigma(1)}-x_{\sigma(2)})\dots(x_{\sigma(n-1)}-x_{\sigma(n)})(x_{\sigma(n)}-x)} 
\eeq
\bea
&& P_{n+1}(x;x_1,\dots,x_n)  \cr
&=& (-1)^{n}\Big[Q_{n+1}(x;x_1,\dots,x_n) - \sum_{j=1}^n {1\over x-x_j}\,\Res_{x'\to x_j} Q_{n+1}(x';x_1,\dots,x_n) \Big]
\eea

\ed

\bt
$P_{n+1}(x;x_1,\dots,x_n)$ is a rational fraction of the variable $x$, whose only poles are at the poles of $a(x), b(x)$ or $c(x)$, and if $n\geq 1$, those poles are of degree at most ${\rm max}(\deg a,\deg b,\deg c)-2$.
\et

\proof{It suffices to prove that $P_{n+1}$ has no pole when $x=x_i$. $Q_{n+1}$ has simple poles at those points, and $P_{n+1}$ is constructed precisely by canceling the residues.}

\bt\label{thdeltaQ}

For $n>2$
\bea
 - \delta_{x_{n}} Q_n(x;x_1,\dots,x_{n-1}) 
&=&  (-1)^{n}\,P_{n+1}(x;x_1,\dots,x_{n-1},x_n)  \cr
&& -(-1)^n {\partial\over \partial x_n}\, {W_{n}(x,x_1,\dots,x_{n-1}) + W_{n}(x_n,x_1,\dots,x_{n-1})\over x-x_n} \cr
&& + R_n(x;x_1,\dots,x_{n-1},x_n)
\eea
where $R_n(x;x_1,\dots,x_{n-1},x_n)$ is a rational fraction of $x$ with only simple poles at $x=x_j, j=1,\dots,n-1$.

It follows that for all $n\geq 1$:
\bea
&&  \delta_{x_{n}} P_n(x;x_1,\dots,x_{n-1}) \cr
&=&  P_{n+1}(x;x_1,\dots,x_{n-1},x_n)  
- {\partial\over \partial x_n}\, {W_{n}(x,x_1,\dots,x_{n-1}) + W_{n}(x_n,x_1,\dots,x_{n-1})\over x-x_n} \cr
&& -\delta_{n,2} \,\, {\partial\over \partial x_n}\,  {1\over (x_1-x_n)^2\,(x-x_n)}
\eea
\et

This theorem is quite technical and is proved in appendix \ref{appthdeltaQ}.

\bt\label{thloopeqdet}
For all $n\geq 0$, the correlation functions $W_k$ satisfy the loop equation:
\bea\label{loopeqfromdet}
- P_{n+1}(x;L)&=&  W_{n+2}(x,x,L)
+ \sum_{J\subset L}\,  W_{1+|J|}(x,J) W_{1+n-|J|}(x,L/J) \cr
&& + \sum_{j=1}^n {d\over d x_j}\,\, {W_{n}(x,L/\{x_j\})- W_{n}(L) \over x-x_j}  
\eea
where $L=\{x_1,\dots,x_n\}$.
\et
\proof{
This theorem is rather technical, and the proof is given in appendix \ref{Appproofthloopeqdet}.

The proof consists in proving it for $n=0$ and then apply recursively the loop operator.

}

\bigskip
We call equation \eq{loopeqfromdet} the "loop equation", because it is the same as the loop equation of the 1-matrix model, it is the same which was used to compute recursively the topological expansion of matrix integrals in \cite{eynloop1mat}.
This implies that, if our system depends on some "large parameter" $N$ in such a way that we have a topological expansion (see \cite{ZJDFG}), then all the terms in the expansion are those computed in \cite{eynloop1mat}, i.e. they are the correlation functions of the symplectic invariants of \cite{EOFg}.

\subsection{Topological expansion}
\label{sectopexpgen}

\medskip
{\bf Hypothesis 1:} {\it Topological expansion}.

{\it Let us assume that our differential system ${\cal D}_N(x)$ depends on some large parameter $N$, in such a way that the correlation functions have a so-called {\bf topological expansion}:
\beq
W_n(x_1,\dots,x_n) = \sum_g N^{2-2g-n} W^{(g)}_n(x_1,\dots,x_n)
\eeq
where all $W_n^{(g)}$'s are algebraic functions.
}

\medskip
We emphasize that not all differential systems have that property, and we are making a strong assumption here. However, this assumption holds for many systems which have practical applications in enumerative geometry and integrable systems for instance \cite{BIPZ, ZJDFG}.

\medskip
The existence of such a topological expansion implies that the rational fraction ${1\over N^2}\,\det{({\cal D}(x))}=-{1\over N^2}\,(W_1^2(x)+W_2(x,x))$ has a large $N$ limit:
\beq
\mathop{{\rm lim}}_{N\to\infty} {1\over N^2}\,\det{({\cal D}(x))} = -{\cal E}_\infty(x)
\eeq
and we write:
\beq
Y(x) = -W_1^{(0)}(x) =  -\sqrt{{\cal E}_\infty(x)}
\eeq

${\cal E}_\infty(x)$ is a rational function of $x$, and $Y(x)$ has branchcut singularities at all odd zeroes of  ${\cal E}_\infty(x)$. It is regular at even zeroes, it just vanishes there.
It can be seen easily that the fact that the $W_n$'s satisfy loop equations, implies by recursion, that each $W_n^{(g)}$ is an algebraic function, with possible branchcut  singularities at the odd zeroes, and poles at the even zeroes.

Let us then make a stronger additional hypothesis to the topological expansion property:

\medskip
{\bf Hypothesis 2:} {\it No pole at even zeroes}.

{\it We assume that the $W_n^{(g)}$  have no poles at the even zeroes of ${\cal E}_\infty(x)$.}

\medskip

Again, this assumption does not hold for all systems, however, it holds for many of the most interesting systems concerning enumerative geometry and integrable systems \cite{BIPZ, ZJDFG}.

We can add a 3rd hypothesis. Our algebraic curve $Y^2={\cal E}_\infty(x)$ is an hyperelliptical curve, whose genus is related to the number of odd zeroes. If the genus is $>0$, the Riemann surface $Y(x)$ is not simply connected, it has $2\times {\rm genus}$ non-contractible cycles, and we may impose some further hypothesis on half those cycles, call them ${\cal A}_i$, $i=1,\dots, {\rm genus}$:

\medskip

{\bf Hypothesis 3:} {\it Fixed filling fractions}

{\it We assume that the $W_n^{(g)}$ with $n+g>1$  satisfy:
\beq
\oint_{{\cal A}_i} W_n^{(g)} = 0.
\eeq

}

\medskip
Notice that this hypothesis is automatically fulfilled if the genus is zero, i.e. if $Y(x)$ has only 1-cut.

\medskip
If we make those three hypothesis, we can solve recursively the loop equations, and the loop equations have then a unique solution. That unique solution was first found in \cite{eynloop1mat}, and formalized in \cite{EOFg}. We get:

\bc\label{cordetsymplinv} 
Relation to symplectic invariants

If our differential system $\Psi(x)=\Psi(x,N)$ depends on some "large parameter" $N$, such that
the correlation functions $W_n(x_1,\dots,x_n)$ have a topological expansion of the form $W_n(x_1,\dots,x_n) = \sum_g N^{2-2g-n} W^{(g)}_n(x_1,\dots,x_n)$, and if the coefficients $W_n^{(g)}$'s are meromorphic functions on the limit spectral curve $Y(x)$ with singularities only at branchpoints, and fixed filling fractions (or 1-cut), then the coefficients $W^{(g)}_n(x_1,\dots,x_n)$ in that expansion are obtained from the symplectic invariants defined in \cite{EOFg} for the spectral curve $y=Y(x)$.

\ec

\proof{
The article \cite{eynloop1mat}, precisely consisted in finding the unique solution of loop equations having the topological expansion property  $W_n = \sum_g N^{2-2g-n} W^{(g)}_n$, together with the condition that the $W_n^{(g)}$'s have singularities only at branchpoints and fixed filling fractions. 
In brief, it consisted in computing the only polynomials $P_n^{(g)}$'s compatible with the hypothesis, using Lagrange interpolation formula, and rewriting it in terms of contour integrals and residues on the spectral curve.
And it proved \cite{eynloop1mat, CE}, that the unique solution of loop equations having the topological expansion property, together with this given analytical structure, is given by the "symplectic invariant correlators" of \cite{EOFg} for the spectral curve $y=Y(x) $.

We shall not enter the details here. 
A definition of symplectic invariants correlators is given in def \ref{defSpinvomng} in section \ref{secdefSpinvomng} below.
The example of the Airy system, with spectral curve $Y(x)=\sqrt{x}$ is treated in further details below.
}

\section{Airy kernel}
\label{secAiry}

Let us now apply all the results of the preceding section to the example of the Airy system, which plays a very important role in the universal law of extreme values statistics \cite{TWlaw}, so, let us study it in details.

\bigskip

The Airy function $Ai(x)$ satisfies the Airy equation:
\beq
Ai''(x) = x\, Ai(x).
\eeq
It can be rewritten as a differential system of order $2$ (we introduce the other independent solution $Bi(x)$ called "Bairy"-function):
\beq
{d \over d x} \Psi(x) = {\cal D}(x)\,\Psi(x)
\virg
\Psi(x) = \pmatrix{ Ai(x)& Bi(x)   \cr Ai'(x) & Bi'(x)}
\virg
\det \Psi =1
\eeq
The differential system ${\cal D}(x)$ is:
\beq
{\cal D}(x) = \pmatrix{ 0 & 1 \cr x & 0}
\virg
\Tr {\cal D}  =0
\eeq

The spectral curve is:
\beq
\hat{\cal E}(x,y) = \det(y-{\cal D}(x))=y^2 - x
\eeq
i.e. 
\beq
y=\sqrt{x}
\eeq
This spectral curve has only one branchpoint, located at $x=0$, and there is only one cut $[0,-\infty)$.

Notice that $x$ is not a good variable on the spectral curve. Instead, $z=\sqrt{x}$ is a good uniformizing variable, we  write:
\beq
x(z) = z^2 \virg y(z)=z.
\eeq
\medskip

The corresponding Christoffel-Darboux kernel is the Airy kernel:
\beq
K_{{\rm Airy}}(x_1,x_2) = {Ai(x_1) Bi'(x_2) - Ai'(x_1)Bi(x_2) \over x_1-x_2}
\eeq
This kernel plays an important role in the Tracy-Widom law, and in extreme values statistics \cite{TWlaw}.

The correlation functions are given by def.\ref{defCorrWdetK}:

\beq
W_1(x) = Ai'(x) Bi'(x) - x\, Ai(x)Bi(x) 
\eeq
and for $n\geq 2$:
\beq
{\cal W}_n(x_1,\dots,x_n) = 
"\det"\left( K_{{\rm Airy}}(x_i,x_j) \right)
\eeq
i.e.
\beq
W_n(x_1,\dots,x_n) = 
 - {\delta_{n,2}\over (x_1-x_2)^2} -(-1)^n\, \sum_{\sigma={\rm cyles}} \prod_{i=1}^n K_{{\rm Airy}}(x_{\sigma(i)},x_{\sigma(i+1)}) 
\eeq

\subsection{Topological expansion}

The Airy and Bairy functions have some well known BKW large $x$ expansion (see \cite{}), in some sectors\footnote{Due to the Stokes'phenomenon, the asymptotics may change from one sector to another, near the essential singularity at $\infty$, see \cite{}. The asymptotics we give here, are in the sector $|{\rm Arg}(x)| <{2\pi\over 3}$}, of the form:
\bea
Ai(x) 
&\sim & {\ee{-{2\over 3} x^{3\over 2}}\over \sqrt{2}\,\,x^{1\over 4}}\,\, (1+ \sum_k c_k x^{-3k/2}) \cr
&\sim & {\ee{-{2\over 3} x^{3\over 2}}\over \sqrt{2}\,\,x^{1\over 4}}\,\, (1+ \sum_{k\geq 1} {(-1)^k\, (6k)!\over 2^{6k}\,3^{2k}\, (2k)!\,(3k)!}\,\, x^{-3k/2}) \cr
&\sim & {\ee{-{2\over 3} x^{3\over 2}}\over \sqrt{2}\,\, x^{1\over 4}}\,\, (1- {5\over 3\,\,\, 2^4\,\, x^{3/2}} + \dots) \cr
\eea
and similarly $Bi$ is obtained by changing the sign of the square-root:
\beq
Bi(x) 
\sim {\ee{+{2\over 3} x^{3\over 2}}\over \sqrt{2}\,\,x^{1\over 4}}\,\, (1+ \sum_k (-1)^k\, c_k x^{-3k/2})
\sim {\ee{+{2\over 3} x^{3\over 2}}\over \sqrt{2}\,\,x^{1\over 4}}\,\, (1+ {5\over 3\,\,\, 2^4\,\, x^{3/2}} + \dots)
\eeq

In order to expand the kernel $K$, it is convenient to introduce a scaling variable $N$, and rescale $x\to N^{2\over 3} x$.
The kernel thus has a $1/N$ expansion of the form:
\beq
K(N^{2\over 3}x_1,N^{2\over 3}x_2) = N^{-2\over 3}\,\, \ee{-{2N \over 3}(x_1^{3\over 2} -x_2^{3\over 2})} \,\,\, \sum_{g=0}^\infty N^{-g}\,\, K^{(g)}(x_1,x_2)
\eeq
where
\beq
K^{(0)}(x_1,x_2) = {1\over 2\,\,\,(x_1 x_2)^{1\over 4}}\,\, {1\over \sqrt{x_1}-\sqrt{x_2}}
\eeq
and every $K^{(g)}(x_1,x_2)$ for $g>0$ is an odd homogeneous rational fraction of $z_1=\sqrt{x_1}, z_2=\sqrt{x_2}$, with poles only at $z_1=0$ or $z_2=0$.

This implies that all correlation functions have a $1/N$ expansion (in fact $1/N^2$ expansion by parity):
\beq
W_n(N^{2\over 3}\,x_1,\dots,N^{2\over 3}\,x_n) = N^{-{2n\over 3}}\,\,\sum_{g=0}^\infty N^{2-2g-n}\, W_n^{(g)}(x_1,\dots,x_n)
\eeq
where $W_n^{(g)}(x_1,\dots,x_n)$ is a homogeneous rational fraction of the variables  $z_i=\sqrt{x_i}$ of degree $6-6g-5 n$, and it is easy to see that the only poles can be at $z_i=0$ (except $W_1^{(0)}$ and $W_2^{(0)}$).

For example the 1-point function has the expansion (see \cite{bergereairy}):
\beq
W_1(z^2) = - z +  \sum_{g=1}^\infty {(6g-3)!! \over 3^g\, 2^{5g}\, g!}\,\, z^{1-6g}
\eeq

For example the 2-point function starts as:
\beq
W_2(z_1^2,z_2^2)  = {1\over 4z_1 z_2}\,\, {1\over (z_1+z_2)^2} + \dots
\eeq
and thus:
\beq
W_2^{(0)}(z_1^2,z_2^2)  = {1\over 4z_1 z_2}\,\, {1\over (z_1+z_2)^2} 
\eeq
which can be rewritten:
\beq
W_2^{(0)}(x_1,x_2) dx_1 dx_2 = {dx_1 dx_2\over 4\sqrt{x_1 x_2}}\,\, {1\over (\sqrt{x_1}+\sqrt{x_2})^2} = {dz_1 dz_2\over (z_1-z_2)^2}- {dx_1 dx_2\over (x_1-x_2)^2}
\eeq
where $x(z)=z^2$, and thus $dx(z)=2zdz$.
In other words, $W_2^{(0)}(x_1,x_2) dx_1 dx_2 +{dx_1 dx_2\over (x_1-x_2)^2}=B(z_1,z_2)$ is the Bergman kernel on the spectral curve $y=\sqrt{x}$.

The fact that the function $W_2^{(0)}$ is closely related to the Bergman kernel of its spectral curve is not surprising and fits with the general theory (see \cite{EOFg}).

\subsection{Loop equations and symplectic invariants}

We have seen from theorem \ref{thloopeqdet}, that the $W_n$'s satisfy the loop equations \eq{loopeqfromdet} (where it can be seen from the degrees, that  $\hat{P}_n(x;x_1,\dots,x_n)=x\delta_{n,0}$):
\bea\label{loopeqAiry}
W_{n+2}(x,x,L) + \sum_{J\subset L} W_{1+|J|}(x,J)  W_{1+n-|J|}(x,L/J) \cr
+ \sum_j {d\over d x_j} { W_{n}(x,L/\{j\})- W_{n}(L)\over x-x_j} = x\,\delta_{n,0}
\eea

Moreover the $W_n$'s have a topological expansion:
\beq
W_n(N^{2\over 3}\,x_1,\dots,N^{2\over 3}\,x_n) = N^{-{2n\over 3}}\,\,\sum_{g=0}^\infty N^{2-2g-n}\, W_n^{(g)}(x_1,\dots,x_n)
\eeq
where $W_n^{(g)}(x_1,\dots,x_n)$ is a rational fraction of the $z_i=\sqrt{x_i}$, with poles only at $z_i=0$.

In other words, the Airy system satisfies the 3 asumptions of section \ref{sectopexpgen}, and thus corollary \ref{cordetsymplinv} applies:

\bt\label{thWngspinvAiry}
$W_n^{(g)}(x_1,\dots,x_n)$ are the symplectic invariant correlators of \cite{EOFg} for the spectral curve $y^2=x$.
\et

This result was stated in \cite{EOFg} without a proof, and thus we just provide here the missing proof in \cite{EOFg} In fact in \cite{EOFg} it was stated in the opposite way, i.e. it was claimed that the symplectic invariant correlators for the spectral curve $y^2=x$, were the Airy correlation functions.

\medskip

A summarized definition of symplectic invariants and their correlators is recalled in section \ref{secdefSpinvomng} below.
Let us just give here the examples of some of the first $W_n^{(g)}$'s, which are 
written explicitely in section 10.5. of \cite{EOFg} for the Airy curve:
\beq
W_1^{(1)}(x) = {1\over (2\sqrt{x})^5}
\virg
W_1^{(2)}(x) = {2\,\,\, 9!!\over 2!\, 3^2\, (2\sqrt{x})^{11}}
\virg
W_1^{(3)}(x) = {2^2\,\,\, 15!!\over 3!\, 3^3\, (2\sqrt{x})^{17}}
\eeq
\beq
W_3^{(0)}(x_1,x_2,x_3) = {1\over 2\,\, ( 4 x_1 x_2 x_3)^{3/2}}
\eeq
\beq
W_4^{(0)}(x_1,x_2,x_3,x_4) = {3\over 2^6\,\, (x_1 x_2 x_3 x_4)^{3/2}}\,\, \left( {1\over x_1}+{1\over x_2}+{1\over x_3}+{1\over x_4}\right)
\eeq
and it was claimed in \cite{EOFg} that:
\beq
W_n^{(0)}(x_1,\dots,x_n) = {(n-3)!\over 2^{2n-2}}\,\, {1\over \prod_{j=1}^n x_j^{3/2}}\,\,\, \sum_{|\l|=n-3}\,\, M_\l(1/x_i)\,\prod_{j} {(2\l_j+1) !!\over \l_j !}
\eeq
where $M_\l(z_1,\dots,z_n)={\rm Sym}(z_1^{\l_1}\dots z_{n}^{\l_n})$ is the elementary symmetric monomial function indexed by the partition $\l$. This claim was later proved by Michel Berg\`ere \cite{bergereairy}.

And so on...

\subsection{Exponential formulae}
\label{secairyexp}


Let us define the symmetric primitives $\Phi_n(x_1,\dots,x_n)$ such that $\partial_{x_1}\dots\partial_{x_n} \Phi_n = W_n$, more precisely:
\beq
\Phi_1(x) = -{2\over 3} x^{3\over 2} \,\, + \int_\infty^x (W_1(x')+\sqrt{x'})\, dx'
\eeq
\bea
\Phi_2(x_1,x_2) 
&=& - \ln{(\sqrt{x_1}+\sqrt{x_2})}  \cr
&& + \int_\infty^{x_1} dx'_1 \int_\infty^{x_2} dx'_2\,\,\,  (W_2(x'_1,x'_2)-{1\over 4\,\sqrt{x'_1}\sqrt{x'_2}\,(\sqrt{x'_1}+\sqrt{x'_2})^2}) \cr
\eea
and if $n>2$:
\beq
\Phi_n(x_1,\dots,x_n) = \int_\infty^{x_1} dx'_1\dots \int_\infty^{x_n} dx'_n\,\,\, W_n(x'_1,\dots,x'_n)
\eeq

\medskip

If we integrate \eq{loopeqAiry} with respect to $x_1,\dots,x_n$, and write collectively $L=\{x_1,\dots,x_n\}$, we find:
\bea\label{loopAiPhi}
\partial_{1}\partial_{2}\Phi_{n+2}(x,x,L) + \sum_{J\subset L} \partial_1 \Phi_{1+|J|}(x,J)  \partial_1 \Phi_{1+n-|J|}(x,L/J) \cr
+ \sum_j { \partial_1\Phi_{n}(x,L/\{j\})- \partial_{1}\Phi_{n}(x_j,L/\{j\})\over x-x_j} 
= x\,\delta_{n,0}
\eea
where $\partial_i$ means derivative with respect to the $i^{\rm th}$ variable.

\subsubsection{Exponential formula for the Airy function}

Then we define:
\beq
f(x)  = \sum_l {1\over l!} \partial_1 \Phi_{l+1}(x,x,\dots,x) 
\eeq
where $\partial_1$ means that we take the derivative with respect to the first variable only, and then set all variables $x_i=x$.

Let us compute:
\beq
{d\over dx}f(x) + f(x)^2
\eeq
We have:
\bea
&& {d\over dx}f(x) + f(x)^2 \cr
&=& \sum_l {1\over l!} \partial_1\partial_2 \Phi_{l+2}(x,x,\dots,x) + \sum_l {1\over l!} \partial_1^2 \Phi_{l+1}(x,x,\dots,x) \cr
&& + \sum_{l_1}\sum_{l_2} {1\over l_1 !}{1\over l_2 !} \partial_1 \Phi_{l_1+1}(x,x,\dots,x) \partial_1 \Phi_{l_2+1}(x,x,\dots,x) \cr
&=& \sum_l {1\over l!} \partial_1\partial_2 \Phi_{l+2}(x,x,\dots,x) + \sum_l {1\over l!} \partial_1^2 \Phi_{l+1}(x,x,\dots,x) \cr
&& + \sum_l {1\over l!}\,\sum_{J\subset L}  \partial_1 \Phi_{1+|J|}(x,J) \partial_1 \Phi_{1+l-|J|}(x,L/J) \cr
&=& x
\eea
where we used \eq{loopAiPhi} with $L=\{x,\dots,x\}$.

Therefore $f(x)$ satisfies the Ricatti equation $f'+f^2=x$, which implies that:
\beq
f(x) = {Ai'(x)\over Ai(x)}
\eeq
and this proves that:
\bt\label{thAiryexp} The Airy function satisfies the exponential formula
\beq
\encadremath{
Ai(x) = \ee{\sum_l {1\over l!} \Phi_l(x,\dots,x)} = \ee{\sum_l {1\over l!} \int^x\dots \int^x W_l}
}\eeq
\et
This equality makes sense order by order in the large $x$ expansion, because to any given power in $x$, the sum is finite.

\medskip
Similarly, we find:
\beq
g(x)  = \sum_l {(-1)^l\over l!} \partial_1 \Phi_{l+1}(x,x,\dots,x) 
 = {Bi'(x)\over Bi(x)}
\eeq
i.e.
\beq
Bi(x) = \ee{\sum_l {(-1)^l\over l!} \Phi_l(x,\dots,x)} = \ee{\sum_l {(-1)^l\over l!} \int^x\dots \int^x W_l}
\eeq

\subsubsection{Exponential formula for the Airy kernel}

Then we define:
\beq
U(x,y) = \sum_{l=1}^\infty {1\over l!} \int_y^x\dots \int_y^x W_l 
\eeq
In other words we have:
\bea
U(x,y) =  \sum_{n+m>0} {(-1)^m\over n!\,m!}\,\, \Phi_{n+m}(\overbrace{x,\dots,x}^{n},\overbrace{y,\dots,y}^{m})
\eea
and write:
\beq
U(x,y) = \ln{Ai(x)} + \ln{Bi(y)} +\ln{(g(y)-f(x))} + \ln{h(x,y)}
\eeq
We are now going to determine the function $h(x,y)$.

The loop equation \eq{loopAiPhi} with $L=(\overbrace{x,\dots,x}^{n},\overbrace{y,\dots,y}^{m})$ implies:
\beq
U_{xx} + U_x^2 -  {U_x+U_y\over x-y} = x
\eeq
and
\beq\label{loopeqUyy}
U_{yy} + U_y^2 +  {U_x+U_y\over x-y} = y
\eeq
That implies for $h(x,y)$:
\beq
h_{xx} + 2h_x (f+{f'\over f-g}) = {h_x+h_y\over x-y}
\eeq
\beq\label{loopeqhyy}
h_{yy} + 2h_y (g-{g'\over f-g}) = - \,{h_x+h_y\over x-y}
\eeq

We shall prove that $h(x,y)=1$.
From its very definition, we see that the function $U(x,y)$ has a power series expansion at large $y$, and thus we have:
\beq
h(x,y) = 1 + \sum_{k=1}^\infty {h_k(x)\over y^{k/2}}
\eeq
where the leading coefficient $h_0=1$ is easily obtained from the leading order behaviors of $Ai$ and $Bi$.

Now, assume that there exists $k>0$ such that $h_k\neq 0$, and let $k={\rm min}\{ \td k\, / \,\, h_{\td k}\neq 0\}$, this implies that
\beq
h_{yy} + 2h_y (g-{g'\over f-g}) + {h_x+h_y\over x-y} \sim - {k\, h_k(x)\over y^{k+1\over 2}} (1+O(y^{-1/2}))
\eeq
and thus equation \eq{loopeqhyy} implies that $h_k=0$ which is a contradiction.
Therefore, $\forall k\geq 1$, $h_k=0$, and therefore $h=1$, and we recognize the Airy kernel:
\beq
K(x,y) = {\ee{U(x,y)}\over x-y}
\eeq

and thus:
\bt The Airy kernel satisfies the exponential formula 
\beq
\encadremath{
K(x,y) = {1\over x-y}\,\ee{\sum_{l=1}^\infty {1\over l!} \int_y^x\dots \int_y^x W_l}
}\eeq
\et

\section{Exponential formula, conjecture}

In this section, we conjecture and argue that  the exponential formula:
\beq
K(x,y) \stackrel{?}{=} {1\over x-y}\,\ee{\sum_{l=1}^\infty {1\over l!} \int_y^x\dots \int_y^x W_l}
\eeq
should hold not only for the Airy system, but also for a much larger class of differential systems, namely those which have a topological expansion satisfying the 3 hypothesis of section \ref{sectopexpgen}, and which have a {\bf 1-cut } spectral curve.

\medskip

First, in order for this exponential formula to make sense, we need to know what the infinite sum in the exponential means, i.e. we need a large parameter to make a power series expansion.

This is the case when we have a topological expansion property, i.e. when $W_n$ has a large $N$ expansion of the type:
\beq
W_n(x_1,\dots,x_n) = \sum_{g=0}^\infty N^{2-2g-n} W_n^{(g)}(x_1,\dots,x_n)
\eeq
then, the exponential formula means:
\bea
\td K(x,y) 
&=& \ee{N\int_y^x W_1^{(0)}(x')dx'}\,\, {{1\over 2}\ee{\int_y^x\int_y^x W_2^{(0)}(x_1',x'_2)dx_1'dx_2'}\over x-y}\,\, \exp{}\Big( \cr
&& \sum_{g=0}^\infty\sum_{l\geq 1+2\delta_{g,0}}\,{N^{2-2g-l}\over l!}\,\int_y^x\dots \int_y^x W_l^{(g)} \Big) \cr
&=& \ee{N\int_y^x W_1^{(0)}(x')dx'}\,\, {{1\over 2}\ee{\int_y^x\int_y^x W_2^{(0)}(x_1',x'_2)dx_1'dx_2'}\over x-y}\,\, \Big[ 1 + \cr
&& + {1\over N}\Big( \int_y^x W_1^{(1)} + {1\over 6}\int_y^x\int_y^x \int_y^x W_3^{(0)}\Big) + O(1/N^2) \Big] \cr
\eea
where the last exponential contains only negative powers of $N$ and can be expanded at large $N$.

\subsection{1-cut spectral curves}

From now on, we make the 3 hypothesis of section  \ref{sectopexpgen}, and we assume that we have a {\bf 1-cut } spectral curve.

\subsubsection{Rational parametrization}

Any algebraic curve of genus $0$ can be parametrized with some rational functions of a complex variable, i,e. there exist two rational functions $x(z)$ and $y(z)$ such that $Y(x(z))=y(z)$. 
Here, our equation $Y^2=-{\cal E}_\infty(x)$ is of degree $2$ in $Y$, and it 
implies that the function $x(z)$ is a rational function of degree $2$ of $z$, it has either 2 simple poles, or a double pole. 

In other words, the data of the function $Y(x)$, is equivalent to the data of two rational functions $x(z)$ and $y(z)$, where $x(z)$ is of degree $2$:
\beq
\left\{\begin{array}{l}
x(z)\cr
y(z)
\end{array}\right.
\eeq

\bigskip

$\bullet$ If $x(z)$ has a double pole, we may always reparametrize $z$ such that the double pole is at $\infty$, and the zeroe of $x'$ is at zero, i.e. we may always choose:
\beq
x(z)=z^2+c
\eeq
This is the case for the Airy system.

\medskip

$\bullet$ If $x(z)$ has a two simple poles, we may always reparametrize $z$ such that the simple poles are at $0$ and $\infty$, i.e. we may always choose:
\beq
x(z) = \gamma(z+{1\over z}) + c
\eeq
This is the case for matrix models differential systems.

\subsubsection{Branchpoints}

The branchpoints $a_i$ are defined as the zeroes of $x'(z)$. There are either 1 or 2 branchpoints.

\medskip
$\bullet$ If $x(z)$ has a double pole, $x(z)=z^2+c$, there is only one branchpoint at $z=a=0$.

\medskip

$\bullet$ If $x(z)$ has a two simple poles, $x(z) = \gamma(z+{1\over z}) + c$, there are two branchpoints at $a_1=+1$ and $a_2=-1$.

\subsubsection{Conjugated points}

All hyperelliptical curves have an involution $Y(x)\to -Y(x)$, i.e. for each $z$, there exists a point $\bar z$, such that:
\beq
x(\bar z)=x(z)
\virg
y(\bar z)=-y(z)
\eeq

\medskip
$\bullet$ If $x(z)$ has a double pole, $x(z)=z^2+c$, we have $\bar z=-z$.

\medskip

$\bullet$ If $x(z)$ has a two simple poles, $x(z) = \gamma(z+{1\over z}) + c$, we have $\bar z = 1/z$.

\subsection{Bergman kernel}

The notion of a Bergman kernel exists for all algebraic curves, but for curves of genus $0$, it is particularly simple:

We define the Bergman kernel:
\beq
B(z_1,z_2) = {1\over (z_1-z_2)^2}
\eeq
it is the only rational fraction having a double pole on the diagonal, and which is integrable (it has no residue, and it decreases as $1/z^2$ at $\infty$).

It can be proved \cite{EOFg}, that the function $W_2^{(0)}$ satisfying our 3 hypothesis, is always:
\beq
W_2^{(0)}(x(z_1),x(z_2)) + {1\over (x(z_1)-x(z_2))^2}= {B(z_1,z_2)\over x'(z_1)x'(z_2)} 
\eeq

Let us compute:
\bea
&& \int_{z_2}^{z_1}\int_{z_2}^{z_1}W_2^{(0)}(x(z_1),x(z_2))\,dx(z_1)\,dx(z_2) \cr
&=& \int_{z_2}^{z_1}\int_{z_2}^{z_1} {dz_1\,dz_2\over (z_1-z_2)^2}-{dx(z_1)\,dx(z_2)\over (x(z_1)-x(z_2))^2}  \cr
&=& - 2 \ln{(z_1-z_2)\over x(z_1)-x(z_2)} - \ln{x'(z_1)x'(z_2)} 
\eea
and therefore:
\beq
{1\over x(z_1)-x(z_2)}\,\ee{{1\over 2}\int_{z_2}^{z_1}\int_{z_2}^{z_1}W_2^{(0)}(x(z_1),x(z_2))\,dx(z_1)\,dx(z_2)}
= {1\over (z_1-z_2)\,\sqrt{x'(z_1)x'(z_2)}}
\eeq

\subsection{Definitions of the symplectic invariants}
\label{secdefSpinvomng}

\bd\label{defSpinvomng}
We define (see \cite{EOFg}) the symplectic invariants correlators:
\beq
\om_1^{(0)}(z) = -y(z)
\eeq
\beq
\om_2^{(0)}(z_1,z_2) = B(z_1,z_2)-{x'(z_1)\,x'(z_2)\over (x(z_1)-x(z_2))^2}
\eeq
\beq\label{defomngKrec}
\om_{n+1}^{(g)}(z_0,J) = \sum_i \Res_{z\to a_i} dz\,\,\Kerec(z_0,z) \Big[ \om_{n+2}^{(g-1)}(z,z,J) + \sum_{h=0}^g \sum'_{I\subset J} \om_{1+|I|}^{(h)}(z,I) \om_{1+j-|I|}(z,J/I) \Big]
\eeq
where $J$ is a collective notation for the $n$ variables $J=\{z_1,\dots,z_n\}$,
where $\Sigma_{I\subset J}$ is the sum over subsets of $J$, and $\sum'$ means that we exclude $(h,I)=(0,\emptyset)$ and $(h,I)=(g,J)$, and where $a_i$ are the branch points $x'(a_i)=0$, and the recursion kernel $\Kerec$ is:
\beq
\Kerec(z_0,z) 
= {1 \over 2y(z)\, x'(\bar z)\,\, (z-z_0)}\,\, 
\eeq

\medskip
We also define the "full" correlators, as formal series:
\beq
\om_n(z_1,\dots,z_n) = \sum_{g=0}^\infty N^{2-2g-n}\, \om^{(g)}_n(z_1,\dots,z_n)
\eeq

\ed

Each correlator $\om_n^{(g)}(z_1,\dots,z_n)$ is a rational fucntion of the  $z_i\in{\cal L}$, and they have poles only at the branchpoints $a_i$ (except $\om_1^{(0)}$ and $\om_2^{(0)}$).

We define the scalar correlators $\hat{W}_n^{(g)}$ by dividing by the $dx_i$'s:
\bd
\beq
\hat{W}_n^{(g)}(x(z_1),\dots,x(z_n))  = {\om_n^{(g)}(z_1,\dots,z_n) \over x'(z_1)\dots x'(z_n)}
\eeq
and their "full" resummed version as formal series:
\beq
\hat{W}_n(x_1,\dots,x_n) = \sum_{g=0}^\infty N^{2-2g-n}\, \hat{W}^{(g)}_n(x_1,\dots,x_n)
\eeq

\ed

The corollary \ref{cordetsymplinv} means that:
\bc
\beq
\hat W_n^{(g)}(x_1,\dots,x_n) =  W_n^{(g)}(x_1,\dots,x_n)
\eeq
\ec

\subsection{Exponential formulae}

Let us define the formal Baker-Akhiezer kernel:
\bd\label{defKformel}
\bea\label{eqdefKformel}
\td{K}(z_1,z_2) 
&=& {\ee{-N\int_{z_2}^{z_1} ydx} \over (z_1-z_2)\,\sqrt{x'(z_1)\,x'(z_2)} }  \,\,\,
\exp{\left[\sum'_{g,l} {N^{2-2g-l}\over l!}\, \underbrace{\int^{z_1}_{z_2}\dots \int^{z_1}_{z_2}}_{l}  \om_{l}^{(g)}\right]}
\eea
where $\sum'$ means that we exclude all terms such that $2-2g-l\geq 0$.
This definition makes sense order by order in the $1/N$ expansion: $\ln{\td K}$ is a formal series in $1/N$, whose coefficients are rational functions of $z_1$ and $z_2$.
\ed

This kernel clearly has the property that:
\beq\label{tdKW1}
\sum_g N^{1-2g}\, W^{(g)}_1(x(z)) = \mathop{{\rm lim}}_{z'\to z} \td{K}(z,z')-{1\over x(z)-x(z')} 
\eeq
i.e. it satisfies the determinantal formula for $W_1$.

We conjecture that:
\begin{conjecture}

\beq
\td{K}(z_1,z_2) 
= K(x(z_1),x(z_2))
\eeq

\end{conjecture}

This conjecture is true for the Airy spectral curve, as we have seen in section \ref{secairyexp}.

A way to prove that conjecture would be to prove that:
\beq\label{conjdeltatdK}
-\, \delta_{z_3}\, \td{K}(z_1,z_2) \stackrel{?}{=}  \td{K}(z_1,z_3)\,\td{K}(z_3,z_2)
\eeq
where 
Indeed, by recursively acting with $\delta_{z_j}$ on \eq{tdKW1}, we would generate the determinantal formula for $W_n$.

\smallskip

Let us verify that this conjecture holds to the leading orders in $1/N$.
To leading orders we have:
\bea
\td{K}(z_1,z_2) 
&=& {\ee{-N\int_{z_2}^{z_1} ydx} \over (z_1-z_2)\,\sqrt{x'(z_1)\,x'(z_2)} }  \,\,\,
\Big[ 1 \cr
&& + {1\over N}\left( \int_{z_2}^{z_1} \om_1^{(1)} + {1\over 6}\int_{z_2}^{z_1}\int_{z_2}^{z_1}\int_{z_2}^{z_1} \om_3^{(0)} \right)  + O(1/N^2) \Big]
\eea
therefore (using $x'(z_{n+1})\,\delta_{z_{n+1}}\, \om_n^{(g)}(z_1,\dots,z_n)= {1\over N} \om_{n+1}^{(g)}(z_1,\dots,z_n,z_{n+1})$), we have:
\bea
&& -\,x'(z_3)\, \delta_{z_3}\ln{\td{K}(z_1,z_2) } \cr
&=& - \int_{z_2}^{z_1}\om_2^{(0)}(z',z_3) - {1\over 2N}\,\int_{z_2}^{z_1}\int_{z_2}^{z_1}\om_3^{(0)}(z'_1,z'_2,z_3) + O(1/N^2)  \cr
&=& - \int_{z_2}^{z_1} {dz'\over (z'-z_3)^2} - {1\over 2N}\,\int_{z_2}^{z_1}\int_{z_2}^{z_1}\om_3^{(0)}(z'_1,z'_2,z_3) + O(1/N^2)  \cr
&=& {(z_1-z_2)\over (z_1-z_3)(z_3-z_2)} - {1\over 2N}\,\int_{z_2}^{z_1}\int_{z_2}^{z_1}\om_3^{(0)}(z'_1,z'_2,z_3) + O(1/N^2)  .\cr
\eea
On the other hand we have:
\bea
 {\td{K}(z_1,z_3) \td{K}(z_3,z_2) \over \td{K}(z_1,z_2) } 
&=& {(z_1-z_2)\over (z_1-z_3)(z_3-z_2)\, x'(z_3)}\, \Big[ 1 + \cr
&& + {1\over 6N} \Big( \int_{z_3}^{z_1}\int_{z_3}^{z_1}\int_{z_3}^{z_1}\om_3^{(0)}+\int_{z_2}^{z_3}\int_{z_2}^{z_3}\int_{z_2}^{z_3}\om_3^{(0)} \cr
&& \qquad \quad -\int_{z_2}^{z_1}\int_{z_2}^{z_1}\int_{z_2}^{z_1}\om_3^{(0)} \Big)+O(1/N^2)\Big] .
\eea
The equality \eq{conjdeltatdK} clearly holds to order $N^0$.

To order $1/N$, the equality to prove is thus:
\bea
 {(z_1-z_2)\, \over 6\,(z_1-z_3)(z_3-z_2)}\, 
 \left( \int_{z_3}^{z_1}\int_{z_3}^{z_1}\int_{z_3}^{z_1}\om_3^{(0)}+\int_{z_2}^{z_3}\int_{z_2}^{z_3}\int_{z_2}^{z_3}\om_3^{(0)}-\int_{z_2}^{z_1}\int_{z_2}^{z_1}\int_{z_2}^{z_1}\om_3^{(0)} \right)  \cr
 \stackrel{?}{=} - \,{1\over 2}\,\int_{z_2}^{z_1}\int_{z_2}^{z_1}\om_3^{(0)}(z'_1,z'_2,z_3) \,.\cr
\eea
or in other words:
\bea\label{eqconj1}
&& {(z_1-z_2)\, \over 3\,(z_1-z_3)(z_3-z_2)}\, 
 \left( \int_{z_3}^{z_1}\int_{z_3}^{z_1}\int_{z_3}^{z_1}\om_3^{(0)}+\int_{z_2}^{z_3}\int_{z_2}^{z_3}\int_{z_2}^{z_3}\om_3^{(0)}-\int_{z_2}^{z_1}\int_{z_2}^{z_1}\int_{z_2}^{z_1}\om_3^{(0)} \right)  \cr
&& + \,\int_{z_2}^{z_1}\int_{z_2}^{z_1}\om_3^{(0)}(z'_1,z'_2,z_3) \cr
 &\stackrel{?}{=}& 0
 \eea

For any rational spectral curve, $\om_3^{(0)}$ is given by \eq{defomngKrec}:
\bea
&& \om_3^{(0)}(z_1,z_2,z_3)  \cr
&=& 2\sum_i \Res_{z\to a_i}\, \Kerec(z_1,z)\,\om_2^{(0)}(z,z_2)\om_2^{(0)}(z,z_3)  \cr
&=& \sum_i {1\over y'(a_i)\,\, x''(a_i)}\, {1\over (a_i-z_1)^2(a_i-z_2)^2(a_i-z_3)^2} \cr
\eea
It can also be rewritten:
\beq
 \om_3^{(0)}(z_1,z_2,z_3)  
= \sum_i \Res_{z\to a_i}\, {dz\over y'(z)\,x'(z)}\,\,{1\over (z-z_1)^2\,(z-z_2)^2\,(z-z_3)^2}  
\eeq

Then, let us define:
\beq
S_{z_1,z_2}(z) = \int_{z_2}^{z_1} B(z,z')\,dz' = {1\over z-z_1}-{1\over z-z_2} = {(z_1-z_2)\over (z-z_1)(z-z_2)}
\eeq
We thus have:
\beq
\int_{z_2}^{z_1}\int_{z_2}^{z_1}\int_{z_2}^{z_1}\om_3^{(0)}(z'_1,z'_2,z'_3) = \sum_i \Res_{z\to a_i}\, {(S_{z_1,z_2}(z))^3\over x'(z)\,y'(z)}\,\,dz
\eeq
and
\beq
\int_{z_2}^{z_1}\int_{z_2}^{z_1}\om_3^{(0)}(z'_1,z'_2,z_3) = \sum_i \Res_{z\to a_i}\, {(S_{z_1,z_2}(z))^2\,B(z,z_3)\over x'(z)\,y'(z)}\,\,dz
\eeq

The equality \eq{eqconj1} is thus:
\bea\label{eqconj2}
 && \sum_i \Res_{z\to a_i} {dz\over x'(z)\,y'(z)}\, \Big[
 {(z_1-z_2)\, \over 3\,(z_1-z_3)(z_3-z_2)}\, \Big( (S_{z_1,z_3}(z))^3 + (S_{z_3,z_2}(z))^3 - (S_{z_1,z_2}(z))^3\Big) \cr
 &&  + (S_{z_1,z_2}(z))^2\,B(z,z_3) \Big] \quad
 \stackrel{?}{=} 0
 \eea

The expression inside the bracket is:
\bea
 && {(z_1-z_2)\, \over 3\,(z_1-z_3)(z_3-z_2)}\, \Big( (S_{z_1,z_3}(z))^3 + (S_{z_3,z_2}(z))^3 - (S_{z_1,z_2}(z))^3\Big) \cr
 &&  + (S_{z_1,z_2}(z))^2\,B(z,z_3) \cr
&=& {(z_1-z_2)\, \over 3\,(z_1-z_3)(z_3-z_2)}\, \Big( (S_{z_1,z_3}(z))^3 + (S_{z_3,z_2}(z))^3 - (S_{z_1,z_3}(z)+S_{z_3,z_2}(z))^3\Big) \cr
 &&  + (S_{z_1,z_2}(z))^2\,B(z,z_3) \cr
&=& {(z_1-z_2)\, \over (z_1-z_3)(z_3-z_2)}\, S_{z_1,z_3}(z)\,S_{z_3,z_2}(z)\,S_{z_2,z_1}(z)    + (S_{z_1,z_2}(z))^2\,B(z,z_3) \cr
&=& \,S_{z_2,z_1}(z)\,\Big[
 {(z_1-z_2)\, \over (z_1-z_3)(z_3-z_2)}\, S_{z_1,z_3}(z)\,S_{z_3,z_2}(z)  
   + S_{z_2,z_1}(z)\,B(z,z_3)  \Big ] \cr
&=& \,S_{z_2,z_1}(z)\,\,\Big[
 {(z_1-z_2)\, \over (z_1-z_3)(z_3-z_2)}\, {(z_1-z_3)\over (z-z_1)(z-z_3)}\,{(z_3-z_2)\over (z-z_3)(z-z_2)} \cr
&&   + {(z_2-z_1)\over (z-z_2)(z-z_1)}\,{1\over (z-z_3)^2} \Big ] \cr
&=& \,S_{z_2,z_1}(z)\,dz^2\,\Big[
 {(z_1-z_2)\, \over (z-z_1)(z-z_2)(z-z_3)^2} \cr
&&   + {(z_2-z_1)\over (z-z_2)(z-z_1)\, (z-z_3)^2} \Big ] \cr
&=& 0
\eea
Therefore we have proved that the conjecture holds to order $1/N$.

However, this method is not the good method to prove it to all orders, and we leave it for a further work.

\section{Conclusion}

In this article, we have proved that for any $2\times 2$ differential system, the determinantal correlation functions do obey loop equations.

This is particularly useful if in addition, our differential system has a topological expansion property, because in that case, the solution of loop equations is known.

\medskip

We have thus completed some of the claims in \cite{EOFg}, regarding integrability,
and we have made the link between those two formulations of integrability.

\medskip

Moreover, we have conjectured an exponential formula for the formal kernel defined from the correlation functions of a spectral curve. We have proved this formula for the Airy kernel, and we have proved that it holds to the first two orders in $1/N$. It remains to be proved in the general case.

Also, we have considered only $2\times 2$ systems for simplicity, but it seems that our method could be extended to higher rank systems.

\section*{Acknowledgments}
We would like to thank L. Chekhov, B. Dubrovin, T. Grava, M. Mari\~ no, N. Orantin, for useful and fruitful discussions on this subject. The work of B.E. is partly supported by the Enigma European network MRT-CT-2004-5652, by the ANR project G\'eom\'etrie et int\'egrabilit\'e en physique math\'ematique ANR-05-BLAN-0029-01,  
by the European Science Foundation through the Misgam program,
by the Quebec government with the FQRNT.


\appendix

\bigskip

{\noindent \huge \bf Appendix}

\section{Proof of theorem \ref{thdeltaD} $\delta_y {\cal D}(x)$}\label{appthdeltaD}

{\bf Theorem \ref{thdeltaD}:}
\beq
- \delta_y {\cal D}(x) = {1\over (x-y)^2}(M(x)+M(y)-1) + {1\over x-y} [{\cal D}(x),M(y)]
\eeq

\bigskip

\proof{
We start from
\beq
\vec\phi'(x) = {\cal D}(x)\,\vec\phi(x)
\eeq
and we apply $\delta_y$, i.e.
\bea
\delta_y \vec\phi'(x) &=& d_x \delta_y \vec\phi(x)  \cr
\delta_y ({\cal D}(x) \vec\phi(x)) &=& d_x \delta_y \vec\phi(x)  \cr
(\delta_y {\cal D}(x))\, \vec\phi(x) +  {\cal D}(x) \delta_y\vec\phi(x) &=& d_x \delta_y \vec\phi(x)  \cr
\eea
and therefore:
\bea
\delta_y {\cal D}(x)\, . \vec\phi(x) 
&=& (d_x-{\cal D}(x))\, \delta_y\vec\phi(x) \cr
&=& (d_x-{\cal D}(x))\,{M(y)\,\vec\phi(x)\over x-y} \cr
&=& \left({[M(y),{\cal D}(x)]\over x-y} - {M(y)\over (x-y)^2} \right)\, \vec\phi(x) \cr
\eea

Which means that $\delta_y{\cal D}(x)$ is the matrix in the RHS modulo a matrix which projects $\vec\phi(x)$ to $\vec 0$, and thus, there exists a function $f(x,y)$ such that:
\beq
\delta_y {\cal D}(x) = {[M(y),{\cal D}(x)]\over x-y} - {M(y)\over (x-y)^2} + f(x,y) (1-M(x))
\eeq
Since $\Tr {\cal D}(x)=0$, by taking the trace we find $f(x,y)=1/(x-y)^2$, and thus:
\beq
\delta_y {\cal D}(x) = {[M(y),{\cal D}(x)]\over x-y} + {1-M(x)-M(y)\over (x-y)^2} 
\eeq

}

\section{Proof of theorem \ref{thdeltaQ} $\delta_{x_n} P_n(x;x_1,\dots,x_{n-1})$}\label{appthdeltaQ}

{\bf Theorem \ref{thdeltaQ}:}

{\it 
For $n>2$
\bea
&& - \delta_{x_{n}} Q_n(x;x_1,\dots,x_{n-1}) \cr
&=&  (-1)^{n}\,P_{n+1}(x;x_1,\dots,x_{n-1},x_n)  \cr
&& -(-1)^n {\partial\over \partial x_n}\, {W_{n}(x,x_1,\dots,x_{n-1}) + W_{n}(x_n,x_1,\dots,x_{n-1})\over x-x_n} \cr
&& + R_n(x;x_1,\dots,x_{n-1},x_n)
\eea
where $R_n(x;x_1,\dots,x_{n-1},x_n)$ is a rational fraction of $x$ with only simple poles at $x=x_j, j=1,\dots,n-1$.

It follows that for all $n$:
\bea
&&  \delta_{x_{n}} P_n(x;x_1,\dots,x_{n-1}) \cr
&=&  P_{n+1}(x;x_1,\dots,x_{n-1},x_n)  
- {\partial\over \partial x_n}\, {W_{n}(x,x_1,\dots,x_{n-1}) + W_{n}(x_n,x_1,\dots,x_{n-1})\over x-x_n} \cr
&& -\delta_{n,2} \,\, {\partial\over \partial x_n}\,  {1\over (x_1-x_n)^2\,(x-x_n)}
\eea
}

{\bf Proof:}

Let us start with the case $n>2$, and then we consider the two special cases $n=1$ and $n=2$.

\subsection{Case $n>2$}

\proof{
We write:
\beq
X_{i_1,i_2,\dots,i_k} = (x_{i_1}-x_{i_2})(x_{i_2}-x_{i_3})\dots(x_{i_{k-1}}-x_{i_k})
\eeq
Let us work modulo rational fractions of $x$ having only simple poles at $x=x_1,\dots,x_{n-1}$.
We have:
\bea
&& - \delta_{x_{n}} Q_n(x;x_1,\dots,x_{n-1}) \cr
&=&  \sum_{j=1}^{n-1} \sum_\sigma {1\over (x_j-x_n)}\, {\Tr {\cal D}(x)M(x_{\sigma(1)})\dots [M(x_{\sigma(j)}),M(x_{n})] \dots M(x_{\sigma(n-1)})\over (x-x_{\sigma(1)})\, X_{\sigma(1),\sigma(2),\dots,\sigma(n-1)}\, (x_{\sigma(n-1)}-x)} \cr
&& +  \sum_\sigma {1\over (x-x_n)}\, {\Tr [{\cal D}(x), M(x_{n})]\, M(x_{\sigma(1)})\dots M(x_{\sigma(n-1)})\over (x-x_{\sigma(1)})\, X_{\sigma(1),\sigma(2),\dots,\sigma(n-1)}\, (x_{\sigma(n-1)}-x)} \cr
&& + \sum_\sigma {1\over (x-x_n)^2}\, {\Tr M(x)M(x_{\sigma(1)})\dots M(x_{\sigma(n-1)})\over (x-x_{\sigma(1)})\, X_{\sigma(1),\sigma(2),\dots,\sigma(n-1)}\, (x_{\sigma(n-1)}-x)} \cr
&& + \sum_\sigma {1\over (x-x_n)^2}\, {\Tr M(x_n)M(x_{\sigma(1)})\dots M(x_{\sigma(n-1)})\over (x-x_{\sigma(1)})\, X_{\sigma(1),\sigma(2),\dots,\sigma(n-1)}\, (x_{\sigma(n-1)}-x)} \cr
&& - \sum_\sigma {1\over (x-x_n)^2}\, {\Tr M(x_{\sigma(1)})\dots M(x_{\sigma(n-1)})\over (x-x_{\sigma(1)})\, X_{\sigma(1),\sigma(2),\dots,\sigma(n-1)}\, (x_{\sigma(n-1)}-x)} \cr
\eea
The first 2 lines rearrange into:
\bea
&&  \sum_{j=1}^{n-1} \sum_\sigma {1\over (x_j-x_n)}\, {\Tr {\cal D}(x)M(x_{\sigma(1)})\dots [M(x_{\sigma(j)}),M(x_{n})] \dots M(x_{\sigma(n-1)})\over (x-x_{\sigma(1)})\, X_{\sigma(1),\sigma(2),\dots,\sigma(n-1)}\, (x_{\sigma(n-1)}-x)} \cr
&& +  \sum_\sigma {1\over (x-x_n)}\, {\Tr [{\cal D}(x), M(x_{n})]\, M(x_{\sigma(1)})\dots M(x_{\sigma(n-1)})\over (x-x_{\sigma(1)})\, X_{\sigma(1),\sigma(2),\dots,\sigma(n-1)}\, (x_{\sigma(n-1)}-x)} \cr
&=&  Q_{n+1}(x;x_1,\dots,x_{n-1},x_n)
\eea
In the 4th line, we write that:
\bea
&& {1\over (x-x_n)^2}\,{(x_n-x_{\sigma(1)})(x_{\sigma(n-1)}-x_n)\over (x-x_{\sigma(1)})(x_{\sigma(n-1)}-x)} \cr
&\equiv&  {1\over (x-x_n)^2} - {1\over (x-x_n)(x_n-x_{\sigma(1)})}-{1\over (x-x_n)(x_n-x_{\sigma(n-1)})}
\eea
Therefore we have:
\bea
&& - \delta_{x_{n}} Q_n(x;x_1,\dots,x_{n-1}) \cr
&\equiv&  Q_{n+1}(x;x_1,\dots,x_{n-1},x_n)  \cr
&& - (-1)^n {W_{n}(x,x_1,\dots,x_{n-1}) + W_{n}(x_n,x_1,\dots,x_{n-1})\over (x-x_n)^2} \cr
&& - \sum_\sigma {1\over x-x_n}\, {\Tr M(x_n)M(x_{\sigma(1)})\dots M(x_{\sigma(n-1)})\over (x_n-x_{\sigma(1)})^2\, X_{\sigma(1),\sigma(2),\dots,\sigma(n-1)}\, (x_{\sigma(n-1)}-x_n)}  \cr
&& + \sum_\sigma {1\over x-x_n}\,\, {\Tr M(x_n)M(x_{\sigma(1)})\dots M(x_{\sigma(n-1)})\over (x_n-x_{\sigma(1)})\, X_{\sigma(1),\sigma(2),\dots,\sigma(n-1)}\, (x_{\sigma(n-1)}-x_n)^2} \cr
&& - \sum_\sigma {1\over (x-x_n)^2}\, {\Tr M(x_{\sigma(1)})\dots M(x_{\sigma(n-1)})\over (x-x_{\sigma(1)})\, X_{\sigma(1),\sigma(2),\dots,\sigma(n-1)}\, (x_{\sigma(n-1)}-x)} \cr
\eea
and notice that:
\bea
&& Q_{n+1}(x;x_1,\dots,x_{n-1},x_n) \cr
&\equiv& (-1)^n\,P_{n+1}(x;x_1,\dots,x_{n-1},x_n) + {1\over x-x_n} \Res_{x'\to x_n} Q_{n+1}(x;x_1,\dots,x_{n-1},x_n) \cr
&\equiv& (-1)^n\,P_{n+1}(x;x_1,\dots,x_{n-1},x_n)  \cr
&& + {1\over x-x_n} \Res_{x'\to x_n}  \sum_\sigma   {\Tr {\cal D}(x') M(x_n)\,M(x_{\sigma(1)})  \dots M(x_{\sigma(n-1)})\over (x'-x_n)(x_n-x_{\sigma(1)})\, X_{\sigma(1),\sigma(2),\dots,\sigma(n-1)}\, (x_{\sigma(n-1)}-x')}  \cr
&& + {1\over x-x_n} \Res_{x'\to x_n}  \sum_\sigma   {\Tr  M(x_n){\cal D}(x') \,M(x_{\sigma(1)})  \dots M(x_{\sigma(n-1)})\over (x'-x_{\sigma(1)})\, X_{\sigma(1),\sigma(2),\dots,\sigma(n-1)}\, (x_{\sigma(n-1)}-x')(x_n-x')}  \cr
&\equiv& (-1)^{n}\,P_{n+1}(x;x_1,\dots,x_{n-1},x_n)  \cr
&& +  \sum_\sigma   {\Tr {\cal D}(x_n) M(x_n)\,M(x_{\sigma(1)})  \dots M(x_{\sigma(n-1)})\over (x-x_n)(x_n-x_{\sigma(1)})\, X_{\sigma(1),\sigma(2),\dots,\sigma(n-1)}\, (x_{\sigma(n-1)}-x_n)}  \cr
&& - \sum_\sigma   {\Tr  M(x_n){\cal D}(x_n) \,M(x_{\sigma(1)})  \dots M(x_{\sigma(n-1)})\over (x-x_n)(x_n-x_{\sigma(1)})\, X_{\sigma(1),\sigma(2),\dots,\sigma(n-1)}\, (x_{\sigma(n-1)}-x_n)}  \cr
&\equiv& (-1)^{n}\,P_{n+1}(x;x_1,\dots,x_{n-1},x_n)  \cr
&& +  \sum_\sigma   {\Tr [{\cal D}(x_n), M(x_n)]\,M(x_{\sigma(1)})  \dots M(x_{\sigma(n-1)})\over (x-x_n)(x_n-x_{\sigma(1)})\, X_{\sigma(1),\sigma(2),\dots,\sigma(n-1)}\, (x_{\sigma(n-1)}-x_n)}  \cr
&\equiv& (-1)^{n}\,P_{n+1}(x;x_1,\dots,x_{n-1},x_n)  \cr
&& +  \sum_\sigma   {\Tr  {\partial\over \partial x_n}\,M(x_n)\,M(x_{\sigma(1)})  \dots M(x_{\sigma(n-1)})\over (x-x_n)(x_n-x_{\sigma(1)})\, X_{\sigma(1),\sigma(2),\dots,\sigma(n-1)}\, (x_{\sigma(n-1)}-x_n)}  \cr
&\equiv& (-1)^{n}\,P_{n+1}(x;x_1,\dots,x_{n-1},x_n)  \cr
&& +   {1\over x-x_n}\, {\partial\over \partial x_n}\, \sum_\sigma   {\Tr M(x_n)\,M(x_{\sigma(1)})  \dots M(x_{\sigma(n-1)})\over (x_n-x_{\sigma(1)})\, X_{\sigma(1),\sigma(2),\dots,\sigma(n-1)}\, (x_{\sigma(n-1)}-x_n)}  \cr
&& +  \sum_\sigma   {\Tr M(x_n)\,M(x_{\sigma(1)})  \dots M(x_{\sigma(n-1)})\over (x-x_n)(x_n-x_{\sigma(1)})^2\, X_{\sigma(1),\sigma(2),\dots,\sigma(n-1)}\, (x_{\sigma(n-1)}-x_n)}  \cr
&& -  \sum_\sigma   {\Tr M(x_n)\,M(x_{\sigma(1)})  \dots M(x_{\sigma(n-1)})\over (x-x_n)(x_n-x_{\sigma(1)})\, X_{\sigma(1),\sigma(2),\dots,\sigma(n-1)}\, (x_{\sigma(n-1)}-x_n)^2}  \cr
&\equiv& (-1)^{n}\,P_{n+1}(x;x_1,\dots,x_{n-1},x_n)  \cr
&& -   {(-1)^n\over x-x_n}\, {\partial\over \partial x_n}\, W_n(x_1,\dots,x_n)  \cr
&& +  \sum_\sigma   {\Tr M(x_n)\,M(x_{\sigma(1)})  \dots M(x_{\sigma(n-1)})\over (x-x_n)(x_n-x_{\sigma(1)})^2\, X_{\sigma(1),\sigma(2),\dots,\sigma(n-1)}\, (x_{\sigma(n-1)}-x_n)}  \cr
&& -  \sum_\sigma   {\Tr M(x_n)\,M(x_{\sigma(1)})  \dots M(x_{\sigma(n-1)})\over (x-x_n)(x_n-x_{\sigma(1)})\, X_{\sigma(1),\sigma(2),\dots,\sigma(n-1)}\, (x_{\sigma(n-1)}-x_n)^2}  \cr
\eea

And therefore:
\bea
&& - \delta_{x_{n}} Q_n(x;x_1,\dots,x_{n-1}) \cr
&\equiv&  (-1)^{n}\,P_{n+1}(x;x_1,\dots,x_{n-1},x_n)  \cr
&& -(-1)^n {\partial\over \partial x_n}\, {W_{n}(x,x_1,\dots,x_{n-1}) + W_{n}(x_n,x_1,\dots,x_{n-1})\over x-x_n} \cr
&& - \sum_\sigma {1\over (x-x_n)^2}\, {\Tr M(x_{\sigma(1)})\dots M(x_{\sigma(n-1)})\over (x-x_{\sigma(1)})\, X_{\sigma(1),\sigma(2),\dots,\sigma(n-1)}\, (x_{\sigma(n-1)}-x)} \cr
\eea

Consider the term in the last line:
\bea
&& - \sum_\sigma {1\over (x-x_n)^2}\, {\Tr M(x_{\sigma(1)})\dots M(x_{\sigma(n-1)})\over (x-x_{\sigma(1)})\, X_{\sigma(1),\sigma(2),\dots,\sigma(n-1)}\, (x_{\sigma(n-1)}-x)} \cr
&\equiv& - {\partial \over \partial x_n}\,\left( {1\over x-x_n}\, \sum_\sigma {\Tr M(x_{\sigma(1)})\dots M(x_{\sigma(n-1)})\over (x_n-x_{\sigma(1)})\, X_{\sigma(1),\sigma(2),\dots,\sigma(n-1)}\, (x_{\sigma(n-1)}-x_n)} \right) \cr
\eea

We have:
\bea
&& \sum_\sigma {\Tr M(x_{\sigma(1)})\dots M(x_{\sigma(n-1)})\over (x_n-x_{\sigma(1)})\, X_{\sigma(1),\sigma(2),\dots,\sigma(n-1)}\, (x_{\sigma(n-1)}-x_n)} \cr
&=& \sum_\sigma {\Tr M(x_{\sigma(1)})\dots M(x_{\sigma(n-1)})\over (x_n-x_{\sigma(1)})\, X_{\sigma(1),\sigma(2),\dots,\sigma(n-1)}\, (x_{\sigma(n-1)}-x_{\sigma(1)})} \cr
&& - \sum_\sigma {\Tr M(x_{\sigma(1)})\dots M(x_{\sigma(n-1)})\over (x_n-x_{\sigma(n-1)})\, X_{\sigma(1),\sigma(2),\dots,\sigma(n-1)}\, (x_{\sigma(n-1)}-x_{\sigma(1)})} \cr
\eea
and in the last line, we replace $\sigma=\td\sigma.S$ where $S$ is the shift $i\to i+1$, thus: 
\bea
&& \sum_\sigma {\Tr M(x_{\sigma(1)})\dots M(x_{\sigma(n-1)})\over (x_n-x_{\sigma(1)})\, X_{\sigma(1),\sigma(2),\dots,\sigma(n-1)}\, (x_{\sigma(n-1)}-x_n)} \cr
&=& \sum_\sigma {\Tr M(x_{\sigma(1)})\dots M(x_{\sigma(n-1)})\over (x_n-x_{\sigma(1)})\, X_{\sigma(1),\sigma(2),\dots,\sigma(n-1)}\, (x_{\sigma(n-1)}-x_{\sigma(1)})} \cr
&=& \sum_{\td\sigma} {\Tr M(x_{\td\sigma(1)})\dots M(x_{\td\sigma(n-1)})\over (x_n-x_{\td\sigma(1)})\, X_{\td\sigma(1),\td\sigma(2),\dots,\td\sigma(n-1)}\, (x_{\td\sigma(n-1)}-x_{\td\sigma(1)})} \cr
&=& 0
\eea

And Finally:
\bea
&& - \delta_{x_{n}} Q_n(x;x_1,\dots,x_{n-1}) \cr
&\equiv&  (-1)^{n}\,P_{n+1}(x;x_1,\dots,x_{n-1},x_n) \cr 
&& -(-1)^n {\partial\over \partial x_n}\, {W_{n}(x,x_1,\dots,x_{n-1}) + W_{n}(x_n,x_1,\dots,x_{n-1})\over x-x_n} \cr
\eea

This implies:
\bea
&&  \delta_{x_{n}} P_n(x;x_1,\dots,x_{n-1}) \cr
&=&  P_{n+1}(x;x_1,\dots,x_{n-1},x_n)  \cr
&& - {\partial\over \partial x_n}\, {W_{n}(x,x_1,\dots,x_{n-1}) + W_{n}(x_n,x_1,\dots,x_{n-1})\over x-x_n} \cr
\eea

}

\subsection{Case $n=1$}

\beq
\delta_{y}\, P_1(x) = 
P_2(x;y)  -\,\, {\partial \over \partial y}\,\,{W_1(x)+W_1(y) \over x-y} 
\eeq

\proof{
We start from 
\beq
P_1(x) = - {1\over 2}\,\Tr {\cal D}^2(x)
\eeq
\bea
&& \delta_{y}\, P_1(x) \cr
&=& - \Tr {\cal D}(x)\, \delta_y\,{\cal D}(x)  \cr
&=& {1\over (x-y)^2}\,\Tr {\cal D}(x)\, (M(x)+M(y)-1)  \cr
&=& {1\over (x-y)^2}\,(\Tr {\cal D}(x)M(x) + \Tr {\cal D}(x)\,M(y) )  \cr
&=& {1\over (x-y)^2}\,\Tr ({\cal D}(x)M(x)+\Tr {\cal D}(y)M(y) + (x-y) \Tr {\cal D}'(y)M(y) ) \cr
&& + {1\over (x-y)^2}\,\Tr ({\cal D}(x)-{\cal D}(y)-(x-y){\cal D}'(y))\,M(y)   \cr
&=& {1\over (x-y)^2}\,\Tr \big( {\cal D}(x)M(x)+\Tr {\cal D}(y)M(y) + (x-y) \Tr {\cal D}'(y)M(y)\cr
&& \qquad +{\cal D}(y)M'(y) \,\big)\cr
&& + {1\over (x-y)^2}\,\Tr ({\cal D}(x)-{\cal D}(y)-(x-y){\cal D}'(y))\,M(y)   \cr
&=& {\partial \over \partial y}\,\,{\Tr {\cal D}(x)M(x)+\Tr {\cal D}(y)M(y) \over x-y} \cr
&& + {1\over (x-y)^2}\, \Tr ({\cal D}(x)-{\cal D}(y)-(x-y){\cal D}'(y))\,M(y)   \cr
&=& -\,\, {\partial \over \partial y}\,\,{W_1(x)+W_1(y) \over x-y} + P_2(x;y) \cr
\eea

}

\subsection{Case $n=2$}

\beq
\delta_{y}\, P_2(x;x_1) = P_3(x;x_1,y)
 -\,\, {\partial \over \partial y}\,\,{W_2(x,x_1)+W_2(y,x_1) +{1\over (y-x_1)^2} \over x-y}  
\eeq

\proof{
We define:
\beq
Q_2(x;x_1) =  {\Tr {\cal D}(x)\,M(x_1)\over (x-x_1)^2}
\eeq
We start from
\beq
P_2(x;x_1) = Q_2(x;x_1)- {1\over (x-x_1)^2} \Res_{x'\to x_1} (x'-x_1)\,Q_2(x';x_1) - {1\over x-x_1} \Res_{x'\to x_1} Q_2(x';x_1) 
\eeq
Let us compute $\delta_y Q_2(x;x_1)$, modulo rational fractions of $x$ having only a simple or a double pole at $x=x_1$:
\bea
&& - \delta_y Q_2(x;x_1) \cr
&=& {1\over (x-x_1)^2\,(x-y)^2}\, \Tr (M(x)+M(y)-1)M(x_1) \cr
&& + {1\over (x-x_1)^2\,(x-y)} \, \Tr [{\cal D}(x),M(y)]\,M(x_1) \cr
&& + {1\over (x-x_1)^2\,(x_1-y)}\, \Tr {\cal D}(x) \, [M(x_1),M(y)] \cr
&=& {1\over (x-x_1)^2\,(x-y)^2}\, \Tr M(x)M(x_1) \cr
&& +{1\over (x-x_1)^2\,(x-y)^2}\, \Tr M(y)M(x_1) - {1\over (x-x_1)^2\,(x-y)^2} \cr
&& - {1\over (x-x_1)\,(x_1-y)\,(y-x)}\, \Tr {\cal D}(x) \, [M(x_1),M(y)] \cr
&=& {W_2(x,x_1)\over (x-y)^2}\, + {(y-x_1)^2 W_2(y,x_1) \over (x-x_1)^2\,(x-y)^2}   +{1\over (x-x_1)^2(x-y)^2}  \cr
&& - {1\over (x-x_1)\,(x_1-y)\,(y-x)}\, \Tr {\cal D}(x) \, [M(x_1),M(y)] \cr
&=& {W_2(x,x_1)+W_2(y,x_1)\over (x-y)^2}\, + { W_2(y,x_1) \over (x-x_1)^2} - {2\, W_2(y,x_1) \over (x-x_1)\,(x-y)}   \cr
&& +{1\over (x-x_1)^2(x-y)^2}   + {1\over (x-x_1)\,(x_1-y)\,(x-y)}\, \Tr {\cal D}(x) \, [M(x_1),M(y)] \cr
&\equiv& {W_2(x,x_1)+W_2(y,x_1)\over (x-y)^2}\, - {2\, W_2(y,x_1) \over (y-x_1)\,(x-y)}   +  {\partial \over \partial y}\, {1\over (x-y)\,(x_1-y)^2}  \cr
&& - Q_3(x;x_1,y) \cr
\eea

Notice that:
\bea
&& Q_3(x;x_1,y)  \cr
&\equiv& P_3(x;x_1,y) + {1\over x-y}\,\Res_{x'\to y}\, Q_3(x';x_1,y) \cr
&\equiv& P_3(x;x_1,y) \cr
&& + {1\over x-y}\,\Res_{x'\to y}\,  {1\over (x'-x_1)\,(x_1-y)\,(y-x')}\, \Tr {\cal D}(x') \, [M(x_1),M(y)] \cr
&\equiv& P_3(x;x_1,y) + {1\over x-y}\, {1\over (y-x_1)^2}\, \Tr {\cal D}(y) \, [M(x_1),M(y)] \cr
&\equiv& P_3(x;x_1,y) - {1\over x-y}\, {1\over (y-x_1)^2}\, \Tr [{\cal D}(y),M(y)] \, M(x_1) \cr
&\equiv& P_3(x;x_1,y) - {1\over x-y}\, {1\over (y-x_1)^2}\, {\partial\over \partial y}\, \Tr M(y) \, M(x_1) \cr
&\equiv& P_3(x;x_1,y) - {1\over x-y}\, {1\over (y-x_1)^2}\, {\partial\over \partial y}\, 
\Big( (y-x_1)^2 W_2(y,x_1) + 1  \Big)  \cr
&\equiv& P_3(x;x_1,y) - {1\over x-y}\, {\partial\over \partial y}\,  W_2(y,x_1)   - {2\over (x-y)\, (y-x_1)}\,  W_2(y,x_1)   \cr
\eea
and therefore:
\beq
- \delta_y Q_2(x;x_1) 
\equiv {\partial\over \partial y}\,{W_2(x,x_1)+W_2(y,x_1)+{1\over (y-x_1)^2}\over x-y}\,  - P_3(x;x_1,y) 
\eeq
which implies:
\beq
\delta_y P_2(x;x_1) 
= P_3(x;x_1,y) - {\partial\over \partial y}\,{W_2(x,x_1)+W_2(y,x_1)+{1\over (y-x_1)^2}\over x-y}\,   
\eeq

}

\section{Proof of theorem \ref{thloopeqdet} loop equations}
\label{Appproofthloopeqdet}

{\bf Theorem \ref{thloopeqdet}:}

{\it
For all $n\geq 0$, the correlation functions $W_n$ satisfy the loop equation:
\bea\label{loopeqfromdetap}
0&=&  W_{n+2}(x,x,L)
+ \sum_{J\subset L}\,  W_{1+|J|}(x,J) W_{1+n-|J|}(x,L/J) \cr
&& + \sum_{j=1}^n {d\over d x_j}\,\, {W_{n}(x,L/\{x_j\})- W_{n}(L) \over x-x_j}  + P_{n+1}(x;L)
\eea
where $L=\{x_1,\dots,x_n\}$.

}

\subsection{Case $n=0$}

\proof{

we have:
\bea
W_1(x)
&=& -\Tr {\cal D}(x)M(x) \cr  
&=& -\Tr {\cal D}(x) \vec\phi(x) \vec\psi(x)^t A \cr  
&=& - \vec\psi(x)^t A {\cal D}(x) \vec\phi(x)  \cr  
\eea
and thus:
\bea
W_1(x)^2
&=&  \vec\psi(x)^t A {\cal D}(x) \vec\phi(x)\vec\psi(x)^t A {\cal D}(x) \vec\phi(x)  \cr  
&=&  \Tr  {\cal D}(x) \vec\phi(x)\vec\psi(x)^t A {\cal D}(x) \vec\phi(x)\vec\psi(x)^t A  \cr  
&=&  \Tr  {\cal D}(x) M(x) {\cal D}(x) M(x)  \cr  
\eea
Beside, from \eq{hatW2MM}, we have:
\bea
W_2(x,x) 
&=& -{1\over 2}\, \Tr M'(x)^2 \cr
&=& -{1\over 2}\, \Tr [{\cal D}(x),M(x)]^2 \cr
&=& \Tr {\cal D}(x)^2 M(x)^2 - \Tr {\cal D}(x) M(x) {\cal D}(x) M(x) \cr
\eea
Therefore
\beq
W_2(x,x) + W_1(x)^2 =  \Tr {\cal D}(x)^2 M(x)^2 
\eeq

Moreover we have:
\bea
\Tr {\cal D}^2 M^2 
&=& \Tr {\cal D}^2 M \cr
&=& \Tr {\cal D}^2 \vec\phi\vec\psi^t A \cr
&=&  \vec\psi^t A {\cal D}^2 \vec\phi \cr
&=& - \vec\psi'^t A \vec\phi' \cr
&=& \td\psi'\phi'- \psi'\td\phi'  \cr
&=& (c\psi+d\td\psi)(a\phi+b\td\phi)- (a\psi+b\td\psi)(c\phi+d\td\phi)  \cr
&=& (bc-ad)\, (\psi\td\phi - \td\psi\phi)   \cr
&=& bc-ad   \cr
&=& -\det {\cal D} = {1\over 2}\,\Tr {\cal D}^2
\eea
and finally:
\beq
W_2(x,x) + W_1(x)^2 = -\det {\cal D}(x) = {1\over 2}\,\Tr {\cal D}^2(x) =- P_1(x)
\eeq
}

\subsection{Case $n=1$}

We have:
\beq
W_2(x,x) + W_1(x)^2 =  {1\over 2}\,\Tr {\cal D}(x)^2
\eeq
Apply $\delta_y$ to both sides, that gives:
\beq
W_3(x,x,y) + 2 W_1(x) \Big( W_2(x,y)+{1\over (x-y)^2}\Big) = -P_2(x;y) + {\partial \over \partial y}\, {W_1(x)+W_1(y)\over x-y}
\eeq
which is equivalent to:
\beq
W_3(x,x,y) + 2 W_1(x)  W_2(x,y) = -P_2(x;y) - {\partial \over \partial y}\, {W_1(x)-W_1(y)\over x-y}
\eeq


\begin{thebibliography}{99}
\bibliographystyle{plain}

\bibitem{Ak} G.Akemann,
``Higher genus correlators for the Hermitian matrix model with multiple cuts'',
{\em Nucl. Phys.} {\bf B482} (1996) 403, hep-th/9606004

\bibitem{Amb} G.Akemann and J.Ambj{\o}rn,
``New universal spectral correlators'',
{\em J.Phys.} {\bf A29} (1996) L555--L560, cond-mat/9606129.

\bibitem{AkV} G. Akemann, G. Vernizzi,
 ``Characteristic Polynomials of Complex Random Matrix Models'',
 Nucl.Phys. B660 (2003) 532-556, hep-th/021205.

\bibitem{AkP} G. Akemann, A. Pottier, ``Ratios of characteristic polynomials in complex matrix models'', J.PHYS.A 37 L453 (2004).

\bibitem{BBT} O. Babelon, D. Bernard, M. Talon, {\em Introduction to Classical Integrable Systems }
(Cambridge University Press).


\bibitem{bergere} M C Berg\`ere,
Correlation functions of complex matrix models,
2006 J. Phys. A: Math. Gen. 39 8749-8773.

\bibitem{bergereairy} M C Berg\`ere, unpublished notes.

\bibitem{BEH} M. Bertola, B.Eynard, J. Harnad,
``Duality, Bi-orthogonal Polynomials and Multi-Matrix Models'',
 Commun.Math.Phys. 229 (2002) 73-120.

\bibitem{borsosh} A. Borodin and A. Soshnikov, "Janossy densities I. Determinantal ensembles", J. Stat. Phys 113 611-622 (2003).


\bibitem{BIPZ}
E.~Br\'ezin, C.~Itzykson, G.~Parisi and J.~B.~Zuber,
``Planar Diagrams,''
Commun.\ Math.\ Phys.\  {\bf 59}, 35 (1978).

\bibitem{CE} L.Chekhov, B.Eynard,
``Hermitian matrix model free energy: Feynman graph technique for all genera'',
{\em J. High Energy Phys.} {\bf JHEP03} (2006) 014, hep-th/0504116.

\bibitem{ZJDFG} P. Di Francesco, P. Ginsparg, J. Zinn-Justin,
``2D Gravity and Random Matrices'',
{\em Phys. Rep.} {\bf 254}, 1 (1995).

\bibitem{eynmehta} B. Eynard, M.L. Mehta, "Matrices coupled in a chain: eigenvalue correlations", {\em J. Phys. A: Math. Gen.} {\bf 31} (1998) 4449-4456.

%
%
\bibitem{eynloop1mat}
 B.~Eynard,
``Topological expansion for the 1-hermitian matrix model correlation
functions,''
arXiv:hep-th/0407261.
 

\bibitem{EOFg}
B. Eynard and N. Orantin, ``Invariants of algebraic curves and topological expansion'',
arXiv:math-ph/0702045, Communications in Number Theory and Physics, Vol 1, Number 2, p347-452.




\bibitem{FyoStr} Y. Fyodorov, E. Strahov,
 ``An exact formula for general spectral correlation function of random Hermitian matrices'',
J.Phys.A36:3203-3214,2003, math-ph/0204051.

\bibitem{harnad1} J. Harnad, "Janossy densities, multimatrix spacing distributions and Fredholm resolvents", Int. Math. Res. Not. 48, 2599-2609 (2004).

\bibitem{kostovhirota} I.K. Kostov, "Bilinear functional equations in 2d quantum gravity", \\
hep-th/9602117.


\bibitem{Mehtabook} M.L. Mehta, {\em Random Matrices},2nd edition,
(Academic Press, New York, 1991).

\bibitem{soshnikov} A. Soshnikov, "Janossy densities of coupled random matrices", Commun. Math. Phys. 251, 447-471 (2004).


\bibitem{Szego} G. Szeg\"{o}, ``Orthogonal Polynomials'', {\em AMS}, Providence, 1967.

\bibitem{TWlaw} C. Tracy, H. Widom, "Level-spacing distributions and the Airy kernel",
Comm. Math. Phys. {\bf 159} (1994) 151-174.



%
%

\end{thebibliography}
\end{document}